\documentclass[aps,pre,superscriptaddress,showpacs,amsmath,amssymb,nofootinbibi,reprint,floatfix]{revtex4-1}

\usepackage{hyperref}
\usepackage{graphicx}

\begin{document}
\title{Bivalent Kinetics: Insights from Many Body Physics}

	 \author{Richard E.~Spinney}
	 \affiliation{School  of  Physics, University  of  New  South  Wales  -  Sydney  2052,  Australia}
	 \affiliation{EMBL-Australia  node  in  Single  Molecule  Science, School of Medical Sciences, University  of  New  South  Wales  -  Sydney  2052,  Australia}
	 \author{Lawrence K.~Lee}
	 \affiliation{EMBL-Australia  node  in  Single  Molecule  Science, School of Medical Sciences, University  of  New  South  Wales  -  Sydney  2052,  Australia}
	 \author{Richard G.~Morris}
	 \affiliation{School  of  Physics, University  of  New  South  Wales  -  Sydney  2052,  Australia}
	 \affiliation{EMBL-Australia  node  in  Single  Molecule  Science, School of Medical Sciences, University  of  New  South  Wales  -  Sydney  2052,  Australia}

\begin{abstract}
	Bivalency confers several concentration-dependent phenomena, including avidity, competitive exchange and multi-site competitive exchange. Since these concepts are crucial for a wide variety of topics in cell and molecular biology, their extension, modification and/or re-purposing is also increasingly important for the design and construction of de-novo synthetic systems at the nanoscale.  In this context, we draw upon classical techniques of statistical physics to revisit bivalency, highlighting that receptor site geometry offers a design modality independent of the chemistry of the individual binding interfaces themselves. Recasting the problem in terms of many-body coordination, we explore extended, translationally-invariant chains and lattices of receptor sites.  This not only brings clarity to behaviours associated with simpler motifs, but also enables us to distil core principles for the rational design of concentration-dependent kinetics in synthetic soft-systems, which centre on the notion of geometric frustration.  In doing so, we also reveal the possibility of other tunable spatio-temporal features, such as correlation lengths, mean-squared displacements and percolation-like transitions.
\end{abstract}

\maketitle

\section*{Introduction}
\begin{figure*}[!htp]
\centering
\includegraphics[width=0.98\textwidth]{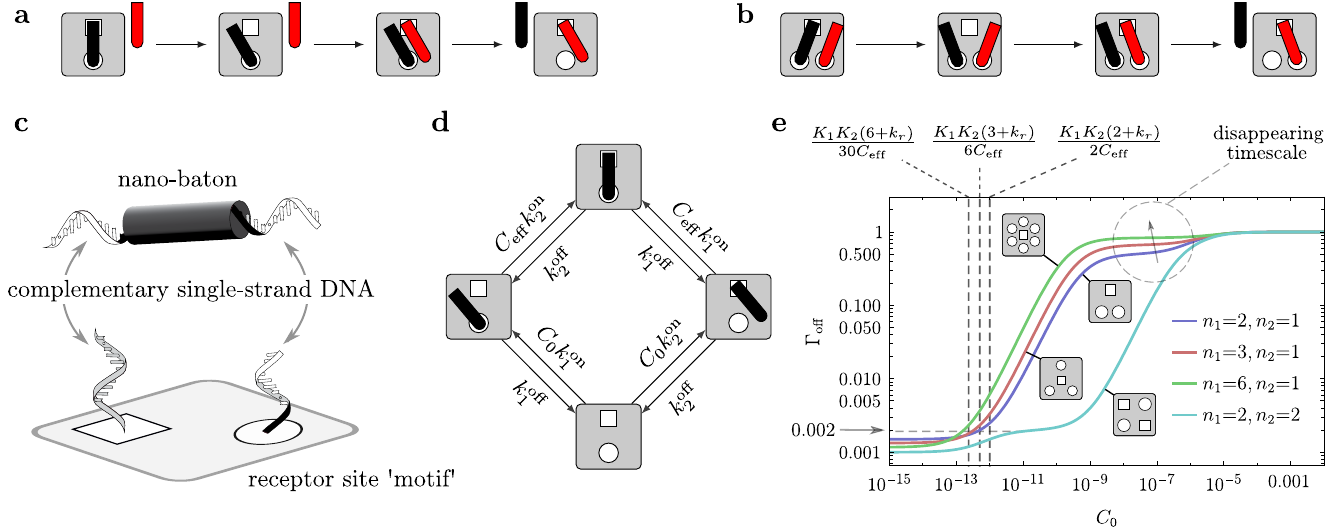}
\caption{
	Bivalent kinetics. Competitive exchange involves a bivalent baton from the bulk (red) occupying a receptor site vacated by the black baton, and therefore destabilising it (panel {\bf a}).  Multiple sites amplify this effect, since the black baton can be destabilised by a red baton that is singly-bound to a neighbouring site, therefore conferring an {\it effective} concentration that is very high (panel {\bf b}). These two mechanisms are detailed quantitatively in Appendix Sec.~\ref{a3}. Nanoscale platforms and batons have recently been engineered via DNA origami such that single strands of DNA at either end of the baton act as tunable binding interfaces whose complementary strands can be arranged in motifs of `primary' (circle) and `secondary' (square) sites (panel {\bf c} and \cite{brownRapidExchangeStably2022}).  A single baton interacting with a primary and secondary receptor site gives rise to a four state system whose transition rates define the principal kinetic parameters outlined in the main text (panel {\bf d}). Mean dissociation rates (per baton) as a function of bulk concentration $C_0$ display two qualitative trends (panel {\bf e}). Vertical dashed lines indicate the characteristic concentration for motifs $n_1=2,3,6$, $n_2=1$ in Eq.~(\ref{char}). Parameters used are $K_2=K_1=10^{-9}\text{M}$, $C_{\rm eff}=10^{-6}\text{M}$, $k_1^{\rm on}=k_2^{\rm on}=10^9 \text{M}^{-1}s^{-1}$, informed by experimental values in \cite{brownRapidExchangeStably2022}.	
	\label{fig1}
}
\end{figure*}

Multivalency underpins several important functions in cell and molecular biology. It simultaneously confers an effective increase in binding affinity \cite{diestlerStatisticalThermodynamicsStability2008,ercolaniAllostericChelateInterannular2011,weberQuantifyingRebindingEffect2012,jencksAttributionAdditivityBinding1981,erlendssonBindingRevisitedAvidity2021,kaneThermodynamicsMultivalentInteractions2010}, so-called {\it avidity}, whilst also facilitating concentration-dependent destabilisation and turnover \cite{zhangControlDNAStrand2009,gibbConcentrationDependentExchangeReplication2014,coccoStochasticRatchetMechanisms2014,singMultiplebindingsiteMechanismExplains2014}, typically referred to as {\it competitive exchange} (Fig.~\ref{fig1}{\bf a}).  These dual mechanisms, alongside the enhancement of competitive exchange due to neighbouring receptor sites--- coined {\it multi-site} competitive exchange \cite{abergStabilityExchangeParadox2016} (Fig.~\ref{fig1}{\bf b})--- underpin a wide variety of diverse phenomena across a range of scales, including toe-hold exchange \cite{zhangControlDNAStrand2009} in DNA hybridisation, liquid-liquid de-mixing \cite{liPhaseTransitionsAssembly2012} and receptor-ligand clustering \cite{conwayMultivalentLigandsControl2013} in sub-cellular aggregates, and specificity \cite{ehrensteinImportanceNaturalIgM2010} in the adaptive immune response of T-cells.

For these reasons, multivalency is also of interest to nanotechnology and synthetic biology \cite{mahonSyntheticMultivalencyBiological2015,huangComingAgeNovo2016,panRecentAdvancesNovo2021}, where expectations surrounding rational design--- {\it i.e.}, the notion that engineering might be informed by an {\it a priori} theoretical or computational characterisation--- form part of the broader narrative \cite{seemanDNANanotechnology2017,ramezaniBuildingMachinesDNA2020}.

In this context, and in conjunction with recent work to engineer and characterise a synthetic DNA-origami system comprising a receptor platform and bivalent `nano-baton' 
(\cite{brownRapidExchangeStably2022} \& Fig.~\ref{fig1}{\bf c}), we now revisit bivalent kinetics, and the response of effective association/dissociation rates to changes in bulk concentration. The perceived challenge here is {\it not} to create individual binding sites with a given affinity, but rather to design {\it motifs} and/or {\it lattices} of receptor sites whose interaction with multivalent entities gives rise to particular kinetic behaviours under changes in bulk concentration.

We are led to introduce extended, translationally-invariant chains and lattices of receptor sites, recasting bi- and therefore multivalent kinetics as a problem of classical many-body coordination and (geometrical) frustration, for which transfer matrices and cavity-like approximations can be brought to bear. This insight permits us to distil core principles that underpin bivalent kinetics, both facilitating rational design and prompting us to speculate on hitherto overlooked areas of biological relevance.

Moreover, we also reveal the possibility of other tunable spatio-temporal features, such as correlation lengths, mean-squared displacements and percolation-like transitions. These latter mechanisms are suggestive of a wider role for multivalency, over and above that of kinetics, and bring to mind several electrically-inspired aims of synthetic soft systems, including switches, circuits, and memory \cite{seemanDNANanotechnology2017}.

\section*{Receptor site geometry as a design modality}
 
Consider a generic bivalent entity--- either nano-baton, protein, or other molecule--- whose binding interfaces are each complementary to one of two types of receptor site (to the exclusion of the other).  We call these receptor sites `primary' and `secondary' following \cite{abergStabilityExchangeParadox2016}, and use the term `baton' for a generic bivalent entity 
(Fig.~\ref{fig1}{\bf c}). In this context, the rate of associations of an unbound baton to individual vacant primary and secondary sites can be written as $C_0k_1^{\rm on}$ and $C_0 k_2^{\rm on}$, respectively, where the bulk concentration of batons is given by $C_0$, and $k_{1/2}^{\rm on}$ are association rates per mole (Fig.~\ref{fig1}{\bf d}). Individual sites are taken to disassociate from their complementary baton ends with rates $k_{1/2}^{\rm off}$, independently of whether the baton is singly or doubly bound. Notably, the rates of association of the unbound ends of singly bound batons are $C_{\rm eff}k_1^{\rm on}$ and $C_{\rm eff}k_2^{\rm on}$, where $C_{\rm eff}$ represents the large {\it effective} concentration \cite{kramerSpanningBindingSites1998,diestlerStatisticalThermodynamicsStability2008,erringtonMechanismsNoncanonicalBinding2019} that arises from the close proximity between receptor sites and unbound baton ends.
\par
A central quantity of interest is the mean dissociation rate of a bound baton, denoted $\Gamma_{\rm off}$, which can be related to (but is not identical to) the effective equilibrium dissociation constant (Appendix Secs.~\ref{a0} \& \ref{a1}). In all practical scenarios, $\Gamma_{\rm off}$ is a monotonic function of the bulk concentration, $C_0$.  As $C_0\to 0$ (depletion), $\Gamma_{\rm off}$ is minimised, and captures `bare' avidity--- {\it i.e.}, stability due to multiple receptor sites with no competitive exchange. As $C_0\to \infty$ (saturation), $\Gamma_{\rm off}$ is maximised, since batons can only bind via one receptor site, and are thus characterised by the nascent dissociation rates of the primary and secondary sites.  Between these two limits, the non-trivial dependence of $\Gamma_\mathrm{off}$ on $C_0$ is dictated  by the geometric arrangement of the receptor sites, which controls the interplay between avidity, competitive exchange and multi-site effects.  As a result, receptor site geometry can be thought of as a configurable design modality that is independent of the chemical or structural properties of the individual receptor sites themselves. 

\section*{All-to-all motifs}

To begin with, we consider motifs with \textit{all-to-all} symmetry.  That is, any pair of primary and secondary sites can be spanned by a single baton (such that all possible binding between the sites forms a complete bipartite graph).  As a result, each motif is uniquely characterised by the number of primary and secondary sites, $n_1$ and $n_2$, respectively.  For a baton with a rigid body, this requires that the distances between all primary and secondary sites are equal. However, this restriction can plausibly be relaxed when the baton is flexible, such that it can associate with two receptor sites over a range of distances.
\par
Notably, a generic  expression for the stationary distribution over baton occupancies can be calculated that encompasses all such all-to-all motifs, from which $\Gamma_{\rm off}$ follows (Appendix Sec.~\ref{a2}) in terms of special functions \cite{10.5555/1098650,50831}. Despite the complicated generic form, a heuristic appreciation of $\Gamma_{\rm off}$ can be obtained from one of only two general cases, outlined in detail in Appendix Sec.~\ref{a4}.
\par
The first case concerns motifs that are one-to-many--- {\it e.g.}, $n_1>1$, $n_2=1$ (Fig.~\ref{fig1}{\bf e}).  Here, as $C_0$ increases, neighbouring sites are increasingly occupied, which facilitates competitive displacement, increasing $\Gamma_{\rm off}$.  The onset of this `multi-site exchange'  (Fig.~\ref{fig1}{\bf b}) depends on the number of neighbouring sites (of opposite type) in the motif. We may identify a characteristic concentration for such an onset (Appendix Sec.~\ref{a4} and Fig.~\ref{fig1}{\bf e}) given as 
\begin{align}
	\frac{C_0^{\rm char}}{C_{\rm eff}}=\frac{K_1K_2(k_r+n_1)}{C^2_{\rm eff}(n_1-1)n_1}+\mathcal{O}(\varepsilon^3), 
	\label{char}
\end{align}
where $K_i=k_i^{\rm off}/k_i^{\rm on}$ are site-specific equilibrium dissociation constants, $k_r=k_2^{\rm on}/k_1^{\rm on}$, and an expansion has been made in the small dimensionless quantity $\varepsilon=K_1/{C_{\rm eff}}$.  On further increases in $C_0$, the multi-site effect plateaus once neighbouring sites are reliably occupied, before giving way to bulk competitive exchange in the traditional sense (Fig.~\ref{fig1}{\bf a}), as $C_0$ approaches (and exceeds) $C_{\rm eff}$.  This secondary stable timescale vanishes for motifs with increasing numbers of neighbours.
\par 
This behaviour can be contrasted with the case $n_1=n_2=2$ (Fig.~\ref{fig1}{\bf e}).  Here, due to the equal site numbers, multi-site effects are effectively eliminated: all partially bound molecules have a complementary site to which they can become doubly bound. As such, the significant increase in $\Gamma_\mathrm{off}$ occurs due to regular competitive exchange from the bulk, whilst the modest increase at low concentrations (rising from a rate of $\sim 0.001$ to $\sim 0.002$ in Fig.~\ref{fig1}{\bf e}) manifests from the removal of vacant neighbouring sites by doubly bound batons, decreasing baton stability as possible rebinding sites become unavailable.

\section*{Chains and Loops}
Context for the aforementioned behaviour is provided by relaxing our requirement of all-to-all symmetry, and replacing it with the weaker constraint of translational symmetry between receptor sites of a given species. This allows us to consider 1D chains of receptor sites that circle back on themselves, forming a loop with $n$ receptor sites. Here, a transfer matrix can be used to solve for $\Gamma_\mathrm{off}$ exactly, for any $n$, so long as primary and secondary sites are equivalent (Appendix Sec.~\ref{a5}).
\par
\begin{figure}[!t]
	\centering
	\includegraphics[width=0.49\textwidth]{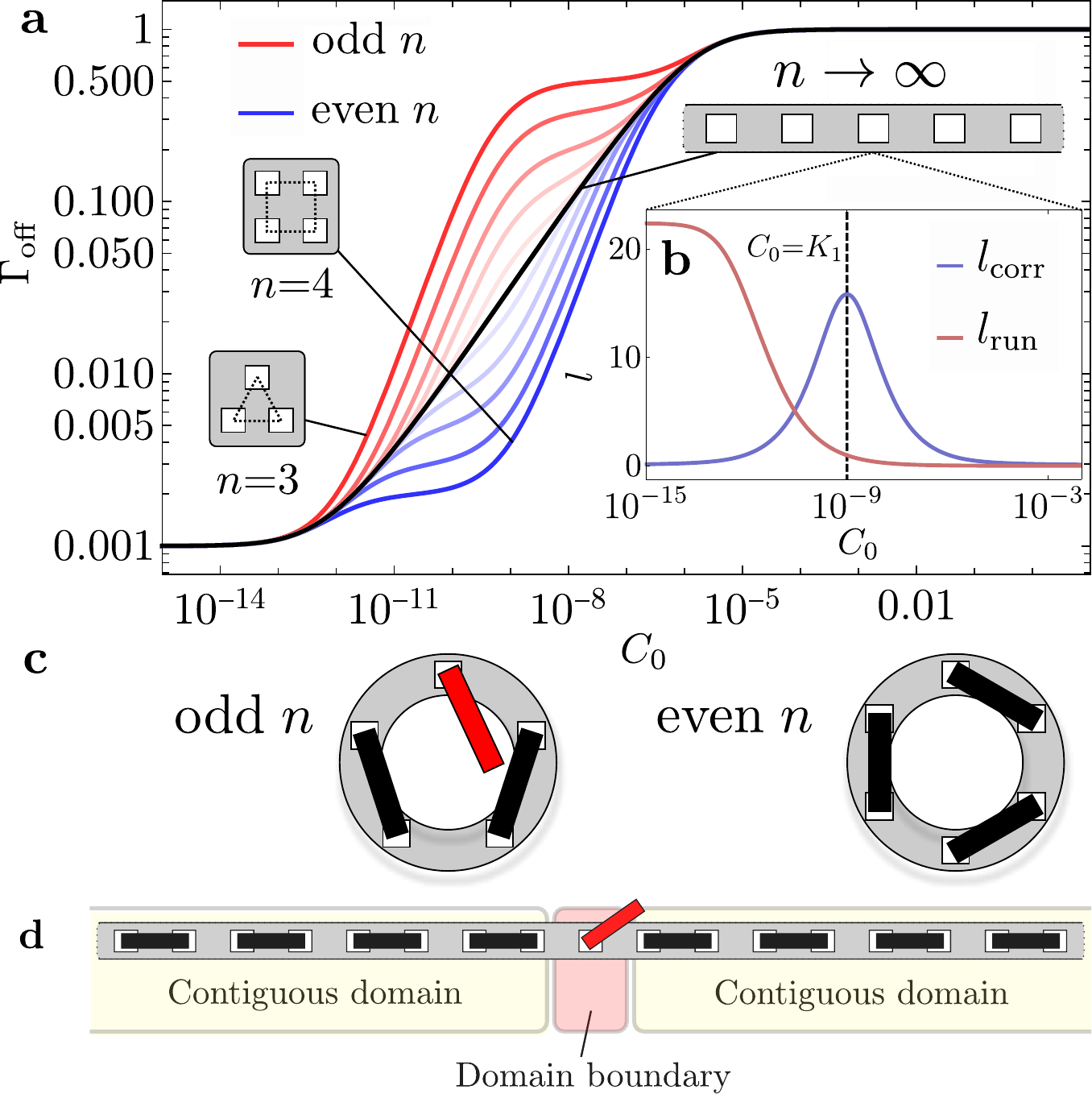}
	\caption{
		Multi-site exchange facilitated by frustrated tiling. Timescales of dissociation for odd (red) and even (blue) loops of length $n$ (panel {\bf a}). Lighter shades indicate higher values, with $n=3,5,9,15,29$ and $n=4,6,10,16,30$, respectively. $n=3$ most closely resembles the $n_1=2$, $n_1=1$ all-to-all motif, whilst $n=4$ can be exactly identified as the $n_1=n_2=2$ all-to-all motif. Both odd and even chains converge on the black line [{\it cf}.~Eq.~(\ref{Goffchain})] as $n\to \infty$, which interpolates between the two main qualitative behaviours (see main text). Run length (approximate) and correlation length (exact) of a single baton in the ($n\to \infty$) $1$D chain (panel {\bf b}, inset). Odd loops cannot perfectly tile whilst even loops can (panel {\bf c}). Spontaneous dislocations or `domain boundaries' provide sites for multi-site exchange (panel {\bf d}).   Parameters used are $K_2=K_1=10^{-9}\text{M}$, $C_{\rm eff}=10^{-6}\text{M}$, $k_1^{\rm on}=k_2^{\rm on}=10^9 \text{M}^{-1}s^{-1}$.
		\label{chain}
	}
\end{figure}

For decreasing {\it odd} values of $n$, we see behaviour that increasingly reflects the aforementioned many-to-one case (Fig.~\ref{chain}{\bf a}, red \& {\bf c}).  We may understand this as arising from an increasing frustration experienced by a baton which cannot reach a more favourable state due to the inability for bivalent molecules to perfectly tile, thus leaving at least one singly bound baton which can participate in multi-site exchange. In contrast, for decreasing {\it even} $n$, multi-site effects are increasingly arrested as the likelihood of perfectly tiling increases (Fig.~\ref{chain}{\bf a}, blue \& {\bf c}), until it reflects the $n_1=n_2=2$ motif for $n=4$.
\par
In the $n\to\infty$ limit, finite size effects decay away such that the parity of $n$ becomes irrelevant (Fig.~\ref{chain}{\bf a}, black), with the role of multi-site exchange being entirely controlled by many-body co-ordination along the lattice, and with energetics of individual batons giving way to entropic contributions of combinations along the chain. The mean dissociation rate in this case takes a particularly simple form (Appendix Sec.~\ref{a5}):
\begin{align}
\Gamma^{n\to\infty}_{\rm off}&=\frac{2C_0K_1k_1^{\rm on}}{C_0-K_1+\eta},
\label{Goffchain}
\end{align}
where $\eta=\sqrt{(C_0+K_1)^2+4C_0C_{\rm eff}}$.
\par
In particular, even when it is favourable for all batons, individually, to be in a perfectly tiled state (such that multi-site effects are absent) the lack of perfect co-ordination over long distances, arising from global entropic contributions, leads to `domain boundaries'--- where contiguous tilings of batons are offset by a single receptor site--- allowing locations for multi-site exchange to occur (Fig.~\ref{chain}{\bf d}). This results in a response to concentration which almost perfectly interpolates between the two extremal behaviours of the all-to-all motifs.  The characteristic correlation length (in units of `sites') associated with chains of doubly-bound batons can be shown to obey
\begin{align}
	l_{\rm corr}^{-1}&=\ln\left[\frac{\eta+(C_0+K_1)}{\eta-(C_0+K_1)}\right],
\end{align}
which is valid for equivalent primary and secondary sites (Fig.~\ref{chain}{\bf b} and Appendix Sec.~\ref{a1}).  This vanishes as $C_0\to 0$ and $C_0\to \infty$, where cross binding is absent and each site is independent, and peaks at $C_0=K_1$ where the system most closely achieves a perfect tiling of cross bound molecules along the chain, with maximum correlation length
\begin{align}
	l^{\rm max}_{\rm corr}&=\frac{C^{1/2}_{\rm eff}}{2K^{1/2}_1}+\frac{K^{1/2}_{1}}{12C^{1/2}_{\rm eff}}+\mathcal{O}(\varepsilon^{\frac{5}{2}}).
\end{align}
For the parameters used in Fig.~\ref{chain}, $l^{\rm max}_{\rm corr}\sim 15$, implying that chains of length $n\gg 15$ are well characterised by the $n\to\infty$ case.

\begin{figure*}[!t]
	\centering
	\includegraphics[width=0.95\textwidth]{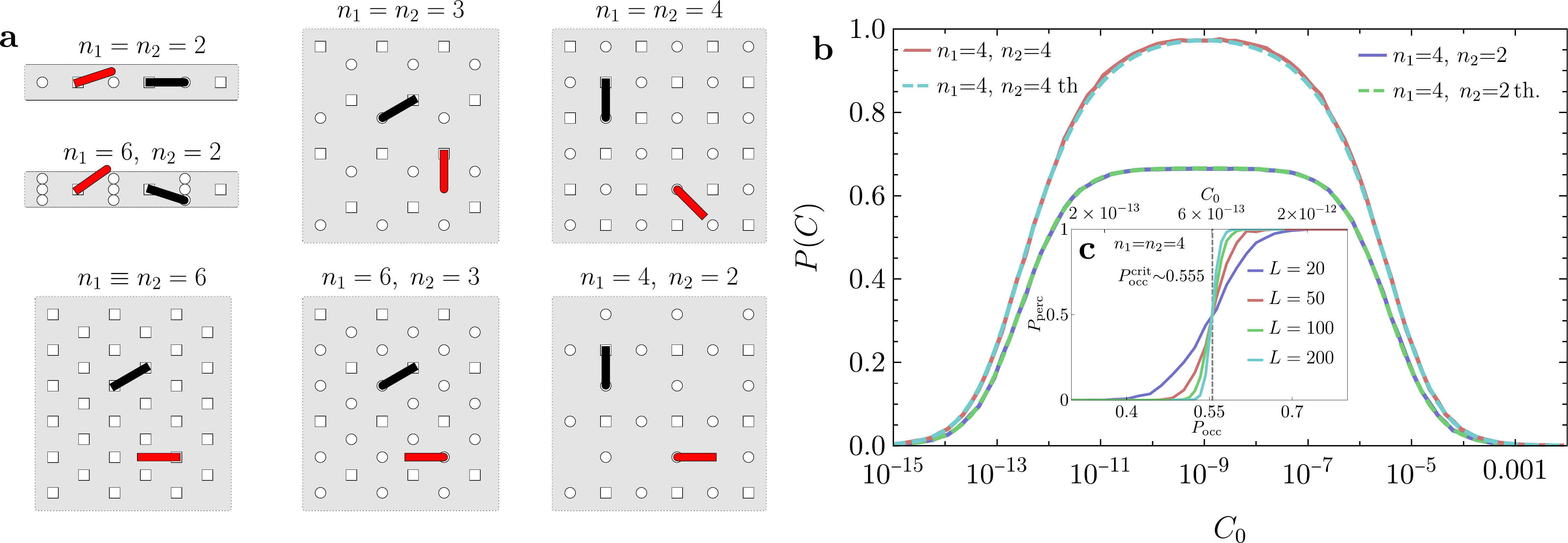}
	\caption{
		Lattices and percolation. Translationally invariant two dimensional lattices with $n_1$ and $n_2$ behaving as co-ordination numbers (panel {\bf a}). The case $n_1=n_2=2$ generalises the $n\to\infty$ $1$D chain. The $n_1=2$, $n_2=6$ case provides an example of an embedding that is only possible if the molecule can bind over a (modest) range of distances, at the expense of strict translational invariance. The case $n_1=n_2=6$ case requires equivalent receptor sites.  A cavity-like approximation [Eq.~(\ref{e12m}) \& dashed curves, panel {\bf b}] agrees with simulation (solid curves, panel {\bf b} - details provided in Appendix Sec.~\ref{a6}). Probability of a percolating cluster against site occupancy and bulk concentration for the $n_1=n_2=4$ lattice where $n=L^2$ (panel {\bf c}).  The critical probability occurs at $P^{\rm crit}_{\rm occ}\simeq 0.555$, below the conventional site percolation  threshold $P^{\rm crit}_{\rm occ}\simeq 0.593$. Parameters used are $K_2=K_1=10^{-9}\text{M}$, $C_{\rm eff}=10^{-6}\text{M}$, $k_1^{\rm on}=k_2^{\rm on}=10^9 \text{M}^{-1}s^{-1}$.
		\label{lattices}
	}
\end{figure*}
More generally, the notion of receptor motifs with (site) translational invariance--- {\it i.e.}, where sites of each type are indistinguishable from each other--- opens the door for a broader class of systems whose spatial embedding simultaneously permits novel kinetics and sophisticated many-body effects.  For example, one can imagine non-trivial \emph{dynamical} behaviours since, once captured, batons can clearly perform a form of diffusive transport by `walking' along the lattice.  In particular this constitutes a realisable system which bears resemblance to both lattice exclusion processes \cite{derridaExactSolutionOnedimensional1992,schutzPhaseTransitionsExactly1993} and stochastic processes with resetting \cite{evansDiffusionStochasticResetting2011,palFirstPassageRestart2017,palFirstPassageRestart2019,evansStochasticResettingApplications2020}, since molecules not only interact with each other through physical occlusion, preventing forward motion, but also compete over receptor sites, thus raising the likelihood of return to the bulk when they do interact.
\par
As an example, one can approximate the mean run length of such motion, implying the existence of a designed, concentration-dependent diffusion constant along the chain. Using a simple combination of the mean life time of a baton and the conditional probability that a neighbour of a given baton is vacant (computed from the transfer matrix), yields
\begin{equation}
	l_{\rm run}^2\simeq\frac{K_1C_{\rm eff}\left(C_0-K_1+\eta\right)}{C_0\left[2K_1(2C_{\rm eff}+K_1)+(C_{\rm eff}+K_1)\left(C_0-K_1+\eta\right)\right]},
\end{equation}
(Fig.~\ref{chain}{\bf b} and Appendix Sec.~\ref{a5}).
\par
\section*{Lattices}
A natural extension of these ideas is to systems of receptor sites with translational symmetry in 2D, where further non-trivial behaviour can be realised. Here, the description in terms of integers $n_1$ and $n_2$ is retained through the interleaving of lattices of primary and secondary sites such that the they become co-ordination numbers--- {\it i.e.}, all primary sites have $n_2$ secondary site neighbours at the baton binding distance, and \emph{vice versa} (Fig.~\ref{lattices}{\bf a}). This description subsumes the one dimensional system (in the limit $n\to\infty$), realised through the choice $n_1=n_2=2$, whilst co-ordination numbers as high as $n_1=n_2=6$ are possible if the primary/secondary binding sites are indistinguishable. Setting either of $n_1$ or $n_2$ to one produces an (infinite number of) all-to-all motif(s).
\par
Exact solutions for lattices with arbitrary coordination numbers are challenging, however we may construct an approximate solution using short range estimates akin to the cavity-method \cite{mezardInformationPhysicsComputation2009}. The approach is detailed in Appendix Sec.~\ref{a6}, but consists of calculating conditional occupation probabilities at a distance of one lattice spacing, whilst neglecting higher order correlations. The central quantity required for computing the kinetics is the expected number of doubly bound batons per $n_1 + n_2$ receptor sites, $\mathbb{E}[N_{c}]$. By defining parameters

\begin{align}
	\gamma &=\frac{{n_1} {n_2} \left({C_0} (C_{\rm eff} ({n_1}+{n_2})+{K_1}+{K_2})+{C_0^2}+{K_1} {K_2}\right)}{2 C_{\rm eff} {C_0} {n_1} {n_2}+2 ({C_0}+{K_1}) ({C_0}+{K_2})},\\
	\beta&=\frac{C_0C_{\rm eff}n_1^2n_2^2}{(C_0+K_1)(C_0+K_2)+C_0C_{\rm eff}n_1n_2},
\end{align}
this can be expressed as
\begin{align}
	\mathbb{E}[N_{c}]&=\gamma-\sqrt{\gamma^2-\beta},
	\label{e12m}
\end{align}
which can be converted to a probability of a random site being occupied by a doubly bound baton, $P(C)=2\mathbb{E}[N_{c}]/(n_1+n_2)$.
\par
For the parameters used in Figs.~\ref{fig1} \& \ref{chain}, this has excellent agreement with simulation, as shown in Fig.~\ref{lattices}{\bf b}. Moreover, such a result is \emph{exact} for both the $n_1>1$, $n_2=1$ all-to-all motif, and for infinite 1D chains where loops are absent. Thus the expression would also be exact for Bethe lattices \cite{baxterExactlySolvedModels1982a}. Notably, in the case of the infinite 1D chain, it allows an exact solution for the case of distinct primary and secondary sites, not available from the transfer matrix approach. 
\par
Qualitatively, the kinetics of such systems follows the principles discerned from our discussion of all-to-all motifs and 1D chains: when co-ordination numbers are equal the system can tile all receptor sites with doubly bound batons, however the ability to do so is restricted by many body co-ordination resulting in the large tunable dissociation rates observed in the $n\to\infty$ 1D chain. In contrast when the co-ordination numbers differ the system becomes frustrated and cannot tile all receptor sites leaving an excess as sites for multi-site exchange resulting in rapid destabilisation at lower concentrations and a fast intermediate timescale.
\par
Many quantitative features of the kinetics, however, depend upon the precise co-ordination numbers of the system. For instance, the dissociation rate at $C_0=0$  is given by
\begin{align}
\Gamma_{\rm off}^{C_0=0}&=\frac{k_1^{\rm on}K_1  K_2 (n_1 + k_r n_2)}{K_2 n_1 +K_1n_2+ C_{\rm eff} n_1 n_2},
\end{align}
allowing much higher stability to be realised on lattices where both $n_1>1$ and $n_2>1$ due to the excess of available receptor sites to all partially bound batons. Other quantitative behaviours involve characteristic destabilisation concentrations which depend, to leading order in $\varepsilon$, on the largest co-ordination number, and timescales of intermediate regimes which depend upon their ratio, both of which are detailed in Appendix Sec.~\ref{a6}.
\par
Moreover, beyond using co-ordination numbers as a modality for rational design of kinetics, such systems also allow for the realisation of more sophisticated many body phenomena. Notably, in two dimensions, the cluster correlation length may diverge as the lattice goes through a percolation transition, where we take a cluster to be group of nearest neighbour occupied sites (either singly or doubly) on the lattice. Here bivalency leads to specific percolation phenomena, very similar (but not identical) to the percolation of pure dimers \cite{cherkasovaPercolationAlignedDimers2010,cornetteDependencePercolationThreshold2003,cornettePercolationPolyatomicSpecies2003,lebrechtAnalyticalApproximationSite2019,leroyerMonteCarloAnalysis1994,tarasevichPercolationLinearMers2012,vandewalleNewUniversalityRandom2000}, where double binding, and the ability to tile, leads to strong neighbour correlations, resulting in a percolation threshold which lies between an estimated lower limit of $P_{\rm occ}\simeq 0.555$ for $K_1\ll C_{\rm eff}$ (valid for Fig.~\ref{lattices}{\bf c}) and the standard result for site percolation with monomers when $C_{\rm eff}\to 0$ ($P_{\rm occ}\simeq 0.593$) \cite{staufferIntroductionPercolationTheory2017}.  Notably, the control parameter for this transition is the bulk concentration, with critical concentration given by $C_0^{\rm crit}/C_{\rm eff}= (4-P_{\rm crit})P_{\rm crit}K_1^2/16(1-P_{\rm crit})^2C^2_{\rm eff}+\mathcal{O}(\varepsilon^2)$ (Appendix Sec.~\ref{a6}).  Therefore, as the transition is approached from below, the diverging correlation length is associated with an increasing degree of competitive exchange.  The implications for the dynamics of the connected domains remains an open question. 

\section*{Discussion}
Arguing that multivalency is best interpreted in the context of classical many-body coordination, we have two main results.

Firstly, the concentration-dependence of bivalent dissociation kinetics can be understood in terms of an overarching heuristic that encompasses all practical receptor site configurations. This hinges on the notion of geometrical frustration: the extent to which a given configuration cannot be perfectly tiled by batons. When perfect tiling is possible ({\it e.g.}, equal coordination numbers) multi-site exchange arises from entropic effects only, and is increasingly subdued as system sizes decrease.  High levels of frustration ({\it e.g.}, highly unequal coordination numbers), by contrast, permit significant multi-site exchange and are largely system-size independent. We note that whilst we have restricted ourselves to examples with explicit symmetries in order to facilitate analytical results, we expect that for a large class of systems these principles will still apply with only limited quantitative deviations arising from finite size, dislocation, or boundary effects.

Secondly, our calculations highlight a fact that has been `hiding in plain sight': bivalency is tantamount to a short range interaction, and hence its effects are synonymous with a variety of emergent spatio-temporal phenomena that rely on many-body coordination.  We choose to focus on correlation lengths, mean squared displacements and percolation, since they allow us to make contact with existing analytical techniques from classical statistical physics.  However, there are undoubtedly more exotic features that might be realised, either by considering higher order multivalency or spatially varying receptor site patterns, for example.

We posit that these ideas may be relevant to sub-cellular scale complexes and molecular machines in biology which, rather than being fixed structures, continually exchange their constituent proteins with the bulk, potentially impacting (and/or facilitating new) function \cite{tuskSubunitExchangeProtein2018}. For example, the error-correcting ability of the DNA replisome has already been linked to binding between a bivalent DNA polymerase and the 6-fold symmetric helicase \cite{abergStabilityExchangeParadox2016,indianiTranslesionDNAPolymerases2009}.  We therefore speculate that the spatio-temporal features associated with closed loops of receptor sites, including correlation lengths, domain boundaries, and the importance of parity, may be relevant for large complexes with rotational symmetry, including the bacterial flagellar motor 
\cite{delalezSignaldependentTurnoverBacterial2010,delalezStoichiometryTurnoverBacterial,yuanAdaptationOutputChemotaxis2012,leleDynamicsMechanosensingBacterial2013} and nuclear pore complex \cite{rabutMappingDynamicOrganization2004,knockenhauerNuclearPoreComplex2016,hakhverdyanDissectingStructuralDynamics2021}.       

More generally, and in the context of recent advances in nano-engineering \cite{brownRapidExchangeStably2022}, we believe that our work paves the way for a wide range of putative soft systems--- electrically-inspired or otherwise \cite{seemanDNANanotechnology2017}--- whose kinetics and emergent spatio-temporal properties might not only be tunable, but designed {\it a priori} in a rational way.

\begin{appendix}
\section{Equilibrium kinetics of multisite receptors}
\label{a0}
Here we describe our general approach for describing the kinetics of systems of multivalent molecules interacting with systems of receptor sites. To do so we define the effective equilibrium dissociation constant $K_{\rm eff}$ of the system of the receptor sites and the relevant association and dissociation inverse timescales/rates $\Gamma_{\rm on}$ and $\Gamma_{\rm off}$ that relate to both molecular turnover and stability, respectively. 
\par
A system of receptor sites at any given bulk concentration will possess an effective binding affinity that relates  the rate of association events of molecules from the bulk and the rate of unbinding of molecules from the sites which quantifies the `strength' with which it binds to the associated molecules. It is important to note that this treatment explicitly distinguishes only between `bound' and `not bound' molecules, not the manner in which molecules are bound. Explicitly, a multivalent molecule bound at a single site is equivalently `bound' as one which is bound at two (or more). It is this additional degree of freedom that allows for the concentration dependent behaviour characteristic of systems with competitive displacement.
\par
This effective affinity can be captured by (the inverse of) the equilibrium dissociation constant, $K_{\rm eff}$ and is defined, for some process $A+B\leftrightharpoons AB$, as the relative concentrations $K_{\rm eff}=[A][B]/[AB]$, where for monovalent species one has total quantities $[A_{\rm tot}]=[A]+[AB]$. For a multivalent receptor system, $[AB]$ is taken as the concentration of bound molecules, whilst $[A]$ and $[B]$ are the concentrations of free molecules and the number of individual unbound receptor sites, respectively. Explicitly, we consider an $n$-site multivalent receptor $R$ with total concentration $[R_{\rm tot}]$ which associates with molecules $M$ with total concentration $[M_{\rm tot}]$. Then, the dissociation constant relates the mean bound concentration $[AB]=[M_b]=\sum_{i=0}^ni[M_iR]$, where $[M_iR]$ is the concentration of multivalent receptors with $i$ bound molecules, to the bulk concentration of molecules $[A]=[M]=[M_{\rm tot}]-[M_b]$ and the number of free receptor sites $[B]=[R_{\rm tot}]\mathbb{E}[n_\emptyset]$, where $\mathbb{E}[n_\emptyset]$ is the mean number of individual vacant receptor sites per system of $n$ sites. Consequently we have
\begin{align}
	K_{\rm eff}&=\frac{[R_{\rm tot}]\mathbb{E}[n_\emptyset]\left([M_{\rm tot}]-[M_b]\right)}{[M_b]}.
\end{align}
We will consider the dilute receptor limit $n[R_{\rm tot}]/[M_{\rm tot}]\ll1$ such that $[M_{\rm tot}]-[M_b]\simeq [M_{\rm tot}]$. In terms of our parameters we thus have $[M_b]= [R_{\rm tot}]\mathbb{E}[N_{\rm b}]$, with $n[R_{\rm tot}]\ll [M_{\rm tot}]$ the total concentration of receptor sites and $\mathbb{E}[N_{\rm b}]$ the expected number of \emph{molecules} associated (either partially or totally) with the system of receptor sites, and $[M_{\rm tot}]=C_0$. We emphasise, $\mathbb{E}[n_\emptyset]$ is a measure of unoccupied sites, whilst $\mathbb{E}[N_b]$ is a measure of bound molecules. For multivalent systems these need not correspond directly to each other - i.e. we have $n\geq {\mathbb{E}[n_\emptyset]}+{\mathbb{E}[N_b]}$, not an equality. This notation, with lower case $n$ representing integer numbers of binding sites, and upper case $N$ representing integer numbers of molecules is used throughout. Consequently we can describe the inverse affinity as
\begin{align}
	K_{\rm eff}&=C_0\frac{\mathbb{E}[n_\emptyset]}{\mathbb{E}[N_b]}.
\end{align}
This can then be contrasted with the natural timescales of the system which maintains a mean bound fraction $\mathbb{E}[N_b]/n\leq 1$. Specifically, we must have that the rate of association and dissociation events must balance in the steady state and so we may introduce
\begin{align}
	n\Gamma_{\rm on}=\mathbb{E}[N_b]\Gamma_{\rm off}
		\label{GoffATA}
\end{align} 
where  $\Gamma_{\rm on}$ is the amortised net association rate \emph{per binding site} and $\Gamma_{\rm off}$ is the amortised dissociation rate \emph{per bound molecule}. $\Gamma_{\rm off}$ behaves as a multivalent analogue to the monovalent quantity $k^{\rm off}$, describing the timescale of dissociation of an individual molecule, thus characterising its stability. However unlike $C_0k^{\rm on}$, $n\Gamma_{\rm on}$ measures the actual number of association events per second, taking into account the reduction caused by receptor sites being occupied. Consequently it is a characterisation of the rate of molecular turnover supported by an average receptor site. For example in the case of a monovalent molecule-receptor system we have $\Gamma_{\rm off}=k^{\rm off}$, but $\Gamma_{\rm on}\neq C_0k^{\rm on}$ (see below).
\par
 Generically, we may find $n\Gamma_{\rm on}$ by calculating probability flux associated with associations from the bulk, proportional to both $C_0$ and the expected number of vacant receptor sites. For systems of $n_1$ and $n_2$ primary and secondary receptor sites we have $n=n_1+n_2$ and we write the expected number of vacant primary and secondary receptor sites are as $\mathbb{E}[n_1^\emptyset]$ and $\mathbb{E}[n_2^\emptyset]$. Consequently the amortised association rate is given by
\begin{align}
	n\Gamma_{\rm on}&=C_0(k_1^{\rm on}\mathbb{E}[n_1^\emptyset]+k_2^{\rm on}\mathbb{E}[n_2^\emptyset]).
	\label{GonATA}
\end{align}
Again, for a monovalent system, we have $n=1$, and $\mathbb{E}[N_b]=1-{\mathbb{E}[n_\emptyset]}=C_0/(C_0+K)$, such that $\Gamma_{\rm on}= C_0k^{\rm on}K/(C_0+K)$, naturally leading to $K_{\rm eff}=k^{\rm off}/k^{\rm on}=K$. Equivalently, wherever $K_{\rm eff}$ is presented in what follows, this limit is achieved by removing the site interaction through setting $C_{\rm eff}=0$ and demanding identical site kinetics, $K_2=K_1$.
\par
As a practical consequence, once $\Gamma_{\rm on}$ is found as per Eq.~(\ref{GonATA}), computing $\mathbb{E}[N_b]$ and $\mathbb{E}[n_\emptyset]$ allows the calculation of both $K_{\rm eff}$ and $\Gamma_{\rm off}$. Note, also, in the case $k_2^{\rm on}=k_1^{\rm on}$, the dissociation rate and effective affinity are simply related as $\Gamma_{\rm off}=k_1^{\rm on}K_{\rm eff}$.

\begin{figure*}[!htp]
	\centering
	\includegraphics[width=0.97\textwidth]{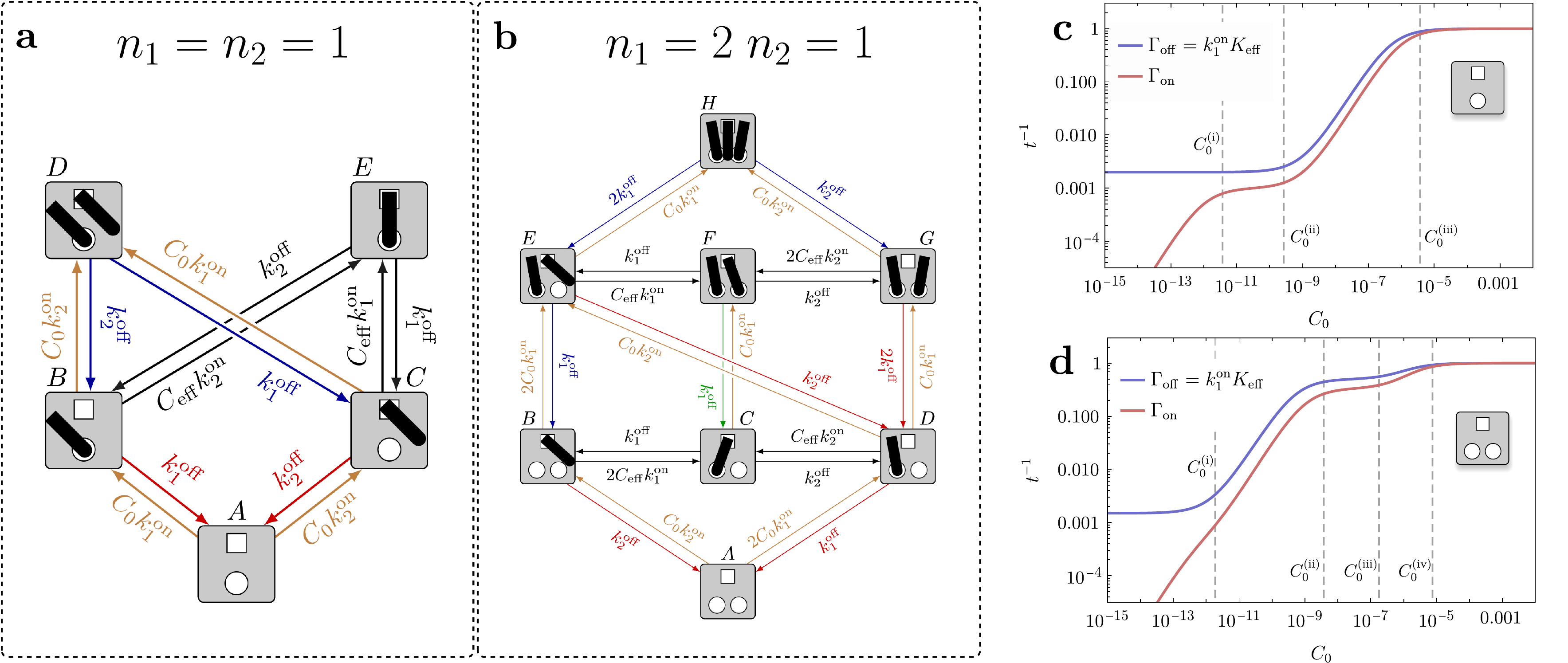}
	\caption{
		State diagrams and allowed transitions for the all-to-all motifs of receptor sites $n_1=n_2=1$ (panel {\bf a}) and $n_1=2$, $n_2=1$ (panel {\bf b}). Brown transitions correspond to associations of a baton (from the bulk) with a previously vacant receptor site. Red transitions indicate a spontaneous dissociation where no competitive exchange was implicated. Blue transitions indicate a dissociation where the molecule was prevented from rebinding due to the presence of a singly bound molecule used as a proxy for canonical competitive exchange. Green transitions indicate a dissociation where the molecule was prevented from rebinding due to the presence of a doubly bound molecule used as a proxy for multi-site competitive exchange. Panels {\bf c} and {\bf d} illustrate kinetic response with concentration. The timescales of the two systems differ significantly. For $n_1=n_2=1$, bound molecules are destabilised by competitive exchange at a concentrations $\sim K_1$, whilst for $n_1=2$, $n_2=1$, they are destabilised at a significantly lower concentrations $\sim K_1^2/C_{\rm eff}$, since the additional site acts as a local reservoir of from which competitive exchange can occur. The multi-site exchange mechanism saturates, however, when such sites are reliably occupied, leading to an additional, intermediate regime that is stable to changes in concentration until bulk concentrations are sufficiently high for canonical competitive exchange to dominate multi-site exchange. Parameters used are $K_2=K_1=10^{-9}\text{M}$, $C_{\rm eff}=10^{-6}\text{M}$, $k_1^{\rm on}=k_2^{\rm on}=10^9 \text{M}^{-1}s^{-1}$. For these values the dissociation constant is merely $K_{\rm eff}=\Gamma_{\rm off}/k_1^{\rm on}$.\label{n11n21}
	}
\end{figure*}
%


\section{Dynamics of nano-batons}
\label{a1}
The set-up presented in the main text is founded upon dynamics of the nano-batons that behave in the following way. We assume that any intra-molecular dynamics that comprise the behaviour of the individual receptor/molecule complexes (e.g. zippering/unzippering) are much faster than the inter-molecular or inter-receptor dynamics, thus allowing us to characterise the receptor site level behaviour through constant rates with associated exponential survival times. As such individual binding/unbinding of receptor site/molecule complexes are assumed to be instantaneous and Markovian with the whole system thus amenable to a master equation formulation
\begin{align}
\dot{P}_i&=\sum_{j\in\mathcal{X}}P_jk_{ji}.
\end{align}
Here $i,j\in\mathcal{X}$ are members of the set of possible binding states $\mathcal{X}$, $P_i$ is the probability of such a state $i$, $k_{ji}$ is the transition rate from state $j$ to state $i$, and $k_{ii}=-\sum_{j\neq i}k_{ij}$ characterises the total escape rate from state $i$. We then treat the system in the steady state where $\dot{P}_i=0\; \forall i\in\mathcal{X}$. The requirement for this steady state to be equilibrium is covered in Sec.~\ref{a2}.
\par
Specific rates follow from the main parameters of the system $k_{1/2}^{\rm on/off}$, $C_{\rm eff}$, and $C_0$, alongside the assumptions that 1) site kinetics are independent of the total binding state of the baton (i.e. if the opposing end is bound or not), ii) that the nano-batons have no preferential orientation when bound at one end and, iii) other than the natural property that only one baton may bind to a given receptor site, the batons otherwise do not interfere with each other. As such if a given receptor site becomes vacated, it is equally likely to be occupied by batons bound at any neighbouring site regardless of the neighbouring site's relative position or whether than baton was the previous occupant of the vacated site. Similarly if there are $m$ singly bound neighbours surrounding a compatible vacant receptor site, the rate at which any of such bound neighbours bind to the vacant site is taken to be $mC_{\rm eff}k_{1/2}^{\rm on}$, i.e. multiply linearly with the number of compatible batons. Analogously, if a singly bound baton is surrounded by $m$ vacant compatible receptor sites, the rate at which it binds to any one of them is also taken to be $mC_{\rm eff}k_{1/2}^{\rm on}$.  
 Examples of such transition behaviour can be observed in Fig.~\ref{n11n21}.
\par
At this point we emphasise: whilst they are important to the quantitative behaviour of the kinetic response of the system, we are not concerned with the precise calculation of the characteristic quantities $K_1$, $K_2$, and $C_{\rm eff}$. Instead we assume these quantities exist, and focus on the behaviour that results from the design of receptor site motifs and lattices. However, it is important to be broadly aware of how these quantities will change with various features of the design of the bivalent molecule. Specifically, in the case of the nano-batons in \cite{brownRapidExchangeStably2022}, the affinities of the receptor sites, characterised by $K_1$ and $K_2$, can primarily be tuned through the sequence and length of the individual DNA strands on either end of the baton, whilst the effective concentration experienced by a vacant receptor site in the presence of a partially bound baton can primarily be tuned through the length of the baton spacer. We do, however, note the use of the separation of scales $K_1\ll C_{\rm eff}$ which motivates the use of a small parameter expansion $\varepsilon=K_1/C_{\rm eff}$ used throughout. Physically, this is required to achieve a doubly bound molecule being substantially thermodynamically favoured over a singly bound molecule. In the main text and this document we use parameters $K_2=K_1=k_1^{\rm off}/k_1^{\rm on}=10^{-9}\text{M}$, $C_{\rm eff}=10^{-6}\text{M}^{-1}$, and $k_1^{\rm on}=k_2^{\rm on}=10^9 \text{M}^{-1}s^{-1}$ informed by experimental values in \cite{brownRapidExchangeStably2022}.
\par
Finally, we highlight that the system, as described, consisting of distinct primary and secondary sites has specific binding behaviour. By this we mean that a primary receptor site can only bind to the associated primary end of the nano-baton, not the secondary end, and \emph{vice versa}. However, there is no reason that batons could not be constructed with identical primary and secondary receptor sites, such that they are of the same species. We refer to this as having `equivalent' receptor sites. Note that this differs from them merely having the same kinetics ($K_2=K_1$, $k_r=k_2^{\rm on}/k_1^{\rm on}=1$). When this is the case this has two main consequences. Firstly, different configurations of receptor sites can fulfil the all-to-all and translational symmetries demanded in this work. For example, the loops considered in Sec.~\ref{a5} do not possess a translational symmetry for odd numbers of receptor sites when the sites are not equivalent. Similarly the $n_1=n_2=6$ lattice described using the method in Sec.~\ref{a6} requires equivalent sites. In contrast, the $n_1=6, n_2=3$ lattice, illustrated in Fig.~3{\bf a} in the main text, requires the specificity that results from distinct receptor site species. The second consequence is that the bulk concentration experienced by any given receptor site effectively doubles since any such site can now be occupied by either end of any given baton in solution, not just one.  As such one can move between descriptions of distinct and equivalent sites (where applicable, notwithstanding the potential loss of symmetry) by simply replacing $C_0\leftrightarrow C_0/2$, as required. The functional forms are otherwise unchanged, and graphically look identical, but shifted, on the log-log graphs that appear in the main text and in this document.


\section{Stationary solutions for all-to-all systems of receptor sites}
\label{a2}

The effective kinetics of a system of receptor sites laid out in Sec. \ref{a0} depends only on the stationary measure of binding configurations through dependence on the number of bound molecules and vacant receptors in that steady state. Here we characterise the steady state distributions and occupancies of systems possessing an all-to-all symmetry, for both equivalent and distinct primary and receptor sites, allowing us to compute the relevant kinetic quantities.
\subsection{Distinct primary and secondary receptor sites}
 Since the molecules and each member of a given species of receptor sites are identical to each other, and with the system possessing an \emph{all-to-all} symmetry by assumption, all binding configurations can be characterised by three integers, $i_1$, $i_2$, and $i_c$, corresponding to the number of singly bound molecules at a primary and secondary site and the number of cross bound molecules, respectively. Crucially, we then insist upon detailed balance, such that the stationary measure must be of the form
\begin{align}
P(i_1,i_2,i_c)&=\frac{h(i_1,i_2,i_c)}{Z}\nonumber\\
&=\frac{g(i_1,i_2,i_c)}{Z}e^{i_c\Delta F_{c}+i_1\Delta F_1+i_2\Delta F_2},
\end{align}
where $g(i_1,i_2,i_c)$ are the number of such possible configurations, $Z$ is the partition sum, and $\Delta F_i$ terms are free energy differences, relative to two vacant receptor sites, multiplied by the Boltzmann factor (in units where $k_BT=1$) which we can identify as
\begin{align}
\Delta F_{c}=\ln\frac{C_0C_{\rm eff}}{K_1K_2},\;\;\Delta F_{1}=\ln\frac{C_0}{K_1},\;\; \Delta F_{2}=\ln\frac{C_0}{K_2}.
\label{f1}
\end{align}
Together with the assumption that the site kinetics are independent of the binding state of the baton, detailed balance is manifest here through receptor sites experiencing an equal effective concentration, $C_{\rm eff}$, regardless of if the partially bound baton is associated at a primary or secondary site, ensuring that the cross-bound state has a well defined free energy. Explicitly there is no allowed notion of having distinct values $C_{\rm eff}^1$ or $C_{\rm eff}^2$, contingent on the initially bound species of receptor site. If this condition was not met a net current of the form bulk $\to$ singly bound at primary $\to$ cross bound $\to$ singly bound at secondary $\to$ bulk would exist, for example.
\par
Given the Boltzmann form, to find the probability of a configuration we therefore need both the configuration function $g$ and the resulting partition sum $Z$. First we specify the configuration function. Starting with an empty set of receptor sites there are $\binom{n_1}{i_1}$ and $\binom{n_2}{i_2}$ combinations of, or ways to place, $i_1$ and $i_2$ singly bound molecules amongst $n_1$ and $n_2$ vacant receptor sites, respectively. Then, in the remaining $n_1-i_1$ and $n_2-i_2$ sites we require the number of combinations for configuring $i_c$ cross bound molecules. This in turn can be expressed as
\begin{align}
&\frac{1}{i_c!}\prod_{j=0}^{i_c-1}(n_1-i_1-j)(n_2-i_2-j)\nonumber\\
&\quad=\frac{(n_1-i_1)!(n_2-i_2)!}{i_c!(n_1-i_1-i_c)!(n_2-i_2-i_c)!}
\end{align}
with the product describing the relevant number of permutations and the factorial converting to a combination such that we are agnostic to ordering. Consequently 
\begin{align}
g(i_1,i_2,i_c)&=\frac{n_1!n_2!}{i_1!i_2!i_c!(n_1-i_1-i_c)!(n_2-i_2-i_c)!}.
\end{align}
The partition sum is then expressible as
\begin{align}
Z=\sum_{i_c=0}^{\text{min}(n_1,n_2)}\sum_{i_1=0}^{n_1-i_c}\sum_{i_2=0}^{n_2-i_c}h(i_1,i_2,i_c).
\end{align}
Assuming, without loss of generality, that $n_1\ge n_2$ we can obtain an expression in terms of special functions
\begin{align}
Z&=(-1)^{n_1}\left(\frac{C_0C_{\rm eff}}{K_1(C_0+K_2)}\right)^{n_1}\left(\frac{C_0+K_2}{K_2}\right)^{n_2}\nonumber\\
&\qquad\times U_{-n_1,1-n_1+n_2,-z}\\
z&=\frac{(C_0{+}K_1)(C_0{+}K_2)}{C_0C_{\rm eff}},
\end{align}
where $U_{-a,b,c}$ is variously described as the Tricomi confluent hypergeometric function, confluent hypergeometric function of the second kind, or simply Kummer confluent hypergeometric function \cite{10.5555/1098650,50831}
\begin{align}
U_{-a,b,c}&=(-1)^{a}\sum_{i=0}^{a}\binom{a}{i}(b+i)_{a-i}(-c)^{i},
\end{align}
and where $(a)_b =\Gamma[a+b]/\Gamma[a]$ is the Pochhammer symbol. 
\par
The expected number of vacant sites, and singly bound molecules can be computed through usual averages of the form
\begin{align}
\mathbb{E}[n_{1}^B]&=\sum_{i_c=0}^{\text{min}(n_1,n_2)}\sum_{i_1=0}^{n_1-i_c}\sum_{i_2=0}^{n_2-i_c}i_1P(i_1,i_2,i_c)\\
\mathbb{E}[n_{1}^\emptyset]&=\sum_{i_c=0}^{\text{min}(n_1,n_2)}\sum_{i_1=0}^{n_1-i_c}\sum_{i_2=0}^{n_2-i_c}(n_1-i_1-i_c)P(i_1,i_2,i_c),
\end{align}
giving the following expressions for number of vacant and single bound molecules at primary and secondary sites, and, as a result, the number of cross bound molecules,
\begin{align}
\mathbb{E}[n_{1}^\emptyset]&=-\frac{n_1K_1(C_0+K_2)}{C_0C_{\rm eff}}\frac{U_{1{-}n_1,2{-}n_1{+}n_2,-z}}{U_{-n_1,1{-}n_1{+}n_2,-z}}\label{ea1}\\
\mathbb{E}[n_{1}^B]&=\frac{C_0}{K_1}\mathbb{E}[n_{1}^\emptyset]\\
\mathbb{E}[n_{2}^\emptyset]&=\frac{n_2K_2}{(C_0+K_2)}\frac{U_{{-}n_1,{-}n_1{+}n_2,-z}}{U_{-n_1,1{-}n_1{+}n_2,-z}}\\
\mathbb{E}[n_{2}^B]&=\frac{C_0}{K_2}\mathbb{E}[n_{2}^\emptyset]\\
\mathbb{E}[N_{c}]&=n_1-\mathbb{E}[n_{1}^\emptyset]-\mathbb{E}[n_{1}^B]=n_2-\mathbb{E}[n_{2}^\emptyset]-\mathbb{E}[n_{2}^B].\label{ean}
\end{align}
One can then compute the association rate per receptor site through Eq.~(\ref{GonATA}) and then, by appreciating that $\mathbb{E}[N_{\rm b}]=\mathbb{E}[n_1^B]+\mathbb{E}[n_2^B]+\mathbb{E}[N_{c}]$, the dissociation rate per molecule can be found through Eq.~(\ref{GoffATA}).
\subsection{Equivalent receptor sites}
We may give a similar description for systems of receptor sites where both primary and secondary sites are equivalent. The resulting binding structure between the sites, assuming an all-to-all symmetry without the distinct specificity, is that of a complete graph (i.e. no longer bipartite). The approach here is identical except that the binding configuration is now specified by only two integers, $i_1$ and $i_c$, and the number of receptor sites is captured with a single integer, $n_1$. Then, once the number of configurations for $i_1$ singly bound molecules to be associated with $n_1$ sites are given as before by $\binom{n_1}{i_1}$, we then require the number of combinations relating to $i_c$ doubly bound molecules associated with $n_1-i_1$ receptor sites. The number of permutations is a repeated falling product of the form $k(k-1)/2\cdot (k-2)(k-3)/2\ldots$, with $k=n_1-i_1$, equal to the repeated multiplication of possible numbers of distinct pairings at each successive placing of  a doubly bound baton on the system. Consequently, the total number of permutations is given by
\begin{align}
\frac{(n_1-i_1)!}{2^{i_c}(n_1-i_1-2i_c)!},
\end{align}
which, as before, is corrected by a factorial so as we are agnostic to the ordering of the batons such that we have
\begin{align}
\frac{(n_1-i_1)!}{2^{i_c}(n_1-i_1-2i_c)!i_c!}.
\end{align}
This then gives configuration function
\begin{align}
g(i_1,i_c)&=\frac{n_1!}{2^{i_c}(n_1-i_1-2i_c)!i_c!i_1!},
\end{align}
such that
\begin{align}
P(i_1,i_c)&=\frac{h(i_1,i_c)}{Z}=\frac{g(i_1,i_c)}{Z}e^{i_c\Delta F_{c}+i_1\Delta F_1},
\end{align}
and where $\Delta F_{c}$ is the same as before, but with $K_2=K_1$.\par
The partition sum is then given by
\begin{align}
Z=\sum_{i_c=0}^{\lfloor n_1/2\rfloor}\sum_{i_1=0}^{n_1-2i_c}h(i_1,i_c),
\end{align}
where $\lfloor x \rfloor$ is the floor function. This gives
\begin{align}
Z&=\begin{cases}\left(-\frac{2C_0C_{\rm eff}}{K_1^2}\right)^{\frac{n_1}{2}}U_{-\frac{n_1}{2},\frac{1}{2},-z},&n_1\;\text{even}\\
\left(-\frac{2C_0C_{\rm eff}}{K_1^2}\right)^{\frac{n_1-1}{2}}\frac{C_0+K_1}{K_1}U_{\frac{1-n_1}{2},\frac{3}{2},-z},&n_1\;\text{odd}.
\end{cases}
\end{align}
$z$ is now given by $(C_0+K_1)^2/C_0C_{\rm eff}$. This in turn leads to
\begin{align}
\mathbb{E}[n_1^B]&=\begin{cases}
-\frac{(C_0+K_1)n_1}{C_{\rm eff}}\frac{U_{\frac{1-n_1}{2},\frac{3}{2},-z}}{U_{-\frac{n_1}{2},\frac{1}{2},-z}},&n_1\;\text{even}\\
-\frac{(C_0+K_1)n_1}{2C_{\rm eff}}\frac{U_{-\frac{n_1}{2},\frac{1}{2},-z}}{U_{\frac{1-n_1}{2},\frac{3}{2},-z}},&n_1\;\text{odd}.
\label{n1equiv}
\end{cases}
\end{align}
We then have $\mathbb{E}[n_1^B]=(C_0/K_1)\mathbb{E}[n_1^\emptyset]$, as before, $\mathbb{E}[N_{c}]=(n_1-\mathbb{E}[n_1^\emptyset]-\mathbb{E}[n_1^B])/2$, and $\mathbb{E}[N_{\rm b}]=\mathbb{E}[n_1^B]+\mathbb{E}[N_{c}]$.
\section{Detailed description of competitive exchange mechanisms through the $n_1=n_2=1$ and $n_1=2, n_2=1$ configurations}
\label{a3}
Here we construct the stationary measure for the system $n_1=n_2=1$ and $n_1=2, n_2=1$ explicitly, examining distinct dissociation fluxes on the graphs given in SI Fig.~\ref{n11n21}, allowing us to characterise concentrations where changes in kinetic behaviour is observed, with specific emphasis on distinct mechanisms of destabilisation, thus characterising the behaviour and hallmarks of both competitive and multi-site exchange. 
\par
The system consisting of a single primary and secondary site $n_1=n_2=1$ can be computed from the continuous Markov dynamics on the graph in the panel {\bf a} of SI Fig.~\ref{n11n21} using the indicated rates. The stationary distribution is proportional to
\begin{align}
|\mathbf{p}^{\rm st}\rangle&\propto K_1K_2|A\rangle+C_0K_2|B\rangle+C_0K_1|C\rangle\nonumber\\
&\quad+C_0^2|D\rangle+C_0C_{\rm eff}|E\rangle.
\end{align}
Rates drawn in brown in SI Fig.~\ref{n11n21} indicate association events allowing us to compute the mean association rate and expected number of associated molecules
\begin{align}
2\Gamma_{\rm on}&=(1+k_r)C_0k_1^{\rm on}\langle \mathbf{p}^{\rm st}|A\rangle\nonumber\\
&\quad+k_rC_0k_1^{\rm on}\langle \mathbf{p}^{\rm st}|B\rangle+C_0k_1^{\rm on}\langle \mathbf{p}^{\rm st}|C\rangle\nonumber\\
&=\frac{C_0k_1^{\rm on}(K_1(C_0+K_2)+k_rK_2(C_0+K_1))}{C_0(C_0+C_{\rm eff}+K_1)+K_2(C_0+K_1)}\\
\mathbb{E}[N_{ b}]&=2\langle\mathbf{p}^{\rm st}|D\rangle+\langle\mathbf{p}^{\rm st}|E\rangle+\langle\mathbf{p}^{\rm st}|B\rangle+\langle\mathbf{p}^{\rm st}|C\rangle\nonumber\\
&=\frac{C_0(2C_0+C_{\rm eff}+K_1+K_2)}{C_0(C_0+C_{\rm eff}+K_1)+K_2(C_0+K_1)},
\end{align}
where we have written $k_2^{\rm on}=k_rk_1^{\rm on}$ leaving $k_1^{\rm on}$ as the sole dimensionful quantity. 
This in turn allows us to write down the effective dissociation constant and per molecule escape rate
\begin{align}
K_{\rm eff}&=\frac{2K_1K_2+C_0(K_1+K_2)}{2C_0+C_{\rm eff}+K_1+K_2}\\
\Gamma_{\rm off}&=\frac{k_1^{\rm on}(K_1(C_0+K_2)+k_rK_2(C_0+K_1))}{2C_0+C_{\rm eff}+K_1+K_2}.
\end{align}
The results for the system consisting of two primary and one secondary site, $n_1=2, n_2=1$, follow analogously from the stationary measure for the graph and rates illustrated in the panel {\bf b} of SI Fig.~\ref{n11n21},
\begin{align}
|\mathbf{p}^{\rm st}\rangle&\propto K^2_1K_2|A\rangle+C_0K^2_1|B\rangle+2C_0K_1C_{\rm eff}|C\rangle\nonumber\\
&\quad+2C_0K_1K_2|D\rangle+2C_0^2K_1|E\rangle+2C_0^2C_{\rm eff}|F\rangle\nonumber\\
&\quad+C_0^2K_2|G\rangle+C_0^3|H\rangle.
\end{align}
Note that this also can be obtained through Eqs.~(\ref{ean}).
\par
Calculating the relevant kinetic quantities yields
\begin{widetext}
\begin{align}
3\Gamma_{\rm on}&=\frac{{C_0} {k_1^{\rm on}} \left(2 {K_1} ({C_0} ({C_0}+{C_{\rm eff}}+{K_1})+{K_2}
   ({C_0}+{K_1}))+{K_2} {k_r} ({C_0}+{K_1})^2\right)}{({C_0}+{K_1})
   ({C_0} ({C_0}+2 {C_{\rm eff}}+{K_1})+{K_2} ({C_0}+{K_1}))}\\
K_{\rm eff}&=\frac{2 {C_0} {K_1} ({C_0}+{C_{\rm eff}}+{K_1})+{K_2} ({C_0}+{K_1})
   ({C_0}+3 {K_1})}{3 {C_0^2}+2 {C_0} (2 ({C_{\rm eff}}+{K_1})+{K_2})+{K_1} (2
   {C_{\rm eff}}+{K_1}+2 {K_2})}\\
   \Gamma_{\rm off}&=\frac{{k_1^{\rm on}} \left(2 {K_1} ({C_0} ({C_0}+{C_{\rm eff}}+{K_1})+{K_2}
   ({C_0}+{K_1}))+{K_2} {k_r} ({C_0}+{K_1})^2\right)}{3 {C_0^2}+2
   {C_0} (2 ({C_{\rm eff}}+{K_1})+{K_2})+{K_1} (2 {C_{\rm eff}}+{K_1}+2 {K_2})}
\end{align}
\end{widetext}

These expressions have relatively simple behaviour in a qualitative sense, but are themselves relatively unwieldy. And, moreover, whilst all the behaviour derives from the stationary measures on the respective graphs, it is instructive to obtain a mechanistic understanding of the regimes of both timescales and affinities that emerge across  different concentrations, allowing us to define characteristic concentrations that separate these distinct behaviours. To do this we utilise the expected occupation alongside the flux associated with distinct mechanisms of dissociation. All dissociations occur when the molecule is in a singly bound state and so the mechanisms are distinguished by the occupancy of the surrounding receptor sites. As such we define a \emph{spontaneous dissociation} as one which occurs whilst there is a vacant receptor site with which the disassociating molecule could reassociate. We then define a \emph{competitive dissociation} as one which occurs whilst the receptor site(s) the molecule would reassociate with are occupied by singly bound molecules from the bulk. Finally, we then define a \emph{multi-site competitive dissociation} as one which occurs when such neighbouring sites are occupied by cross bound molecules. We may identify transitions in the graphs in SI Fig.~\ref{n11n21} which correspond to such dissociation events as those that are marked in red, blue and, green, respectively. Note that neighbour competitive dissociations do not occur in the $n_1=n_2=1$ case. We may thus construct probabilities of each dissociation mechanism by constructing the appropriate ratios of the fluxes. 
\par
\begin{figure*}[!htp]
\centering
\includegraphics[width=0.975\textwidth]{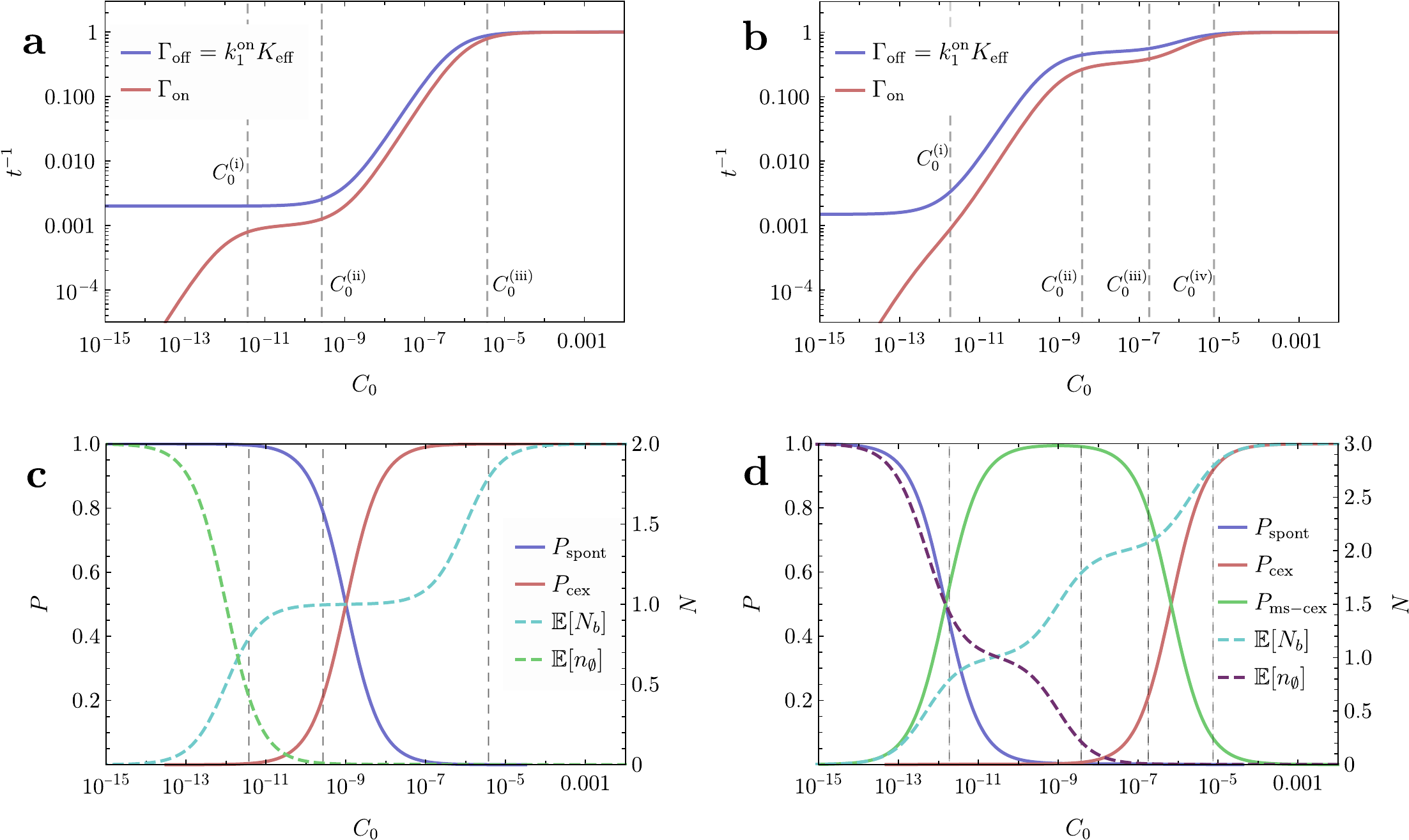}
\caption{Timescales of association and dissociation, mechanisms of dissociation, and mean occupancy \& vacancy, for the $n_1=n_2=1$ system (panels {\bf a} and {\bf c}) and $n_1=2, n_2=1$ system (panels {\bf b} and {\bf d}). Parameters used are $K_2=K_1=10^{-9}\text{M}$, $C_{\rm eff}=10^{-6}\text{M}$, $k_1^{\rm on}=k_2^{\rm on}=10^9 \text{M}^{-1}s^{-1}$. For these parameters $K_{\rm eff}=\Gamma_{\rm off}/k_1^{\rm on}$.\label{fig1results}}
\end{figure*}

First, we consider the $n_1=n_2=1$ case. We broadly observe three characteristic concentrations separating four regimes. The first change in behaviour relates to a slowing of growth of the net association rate coinciding with an increase/decrease in the occupancy and vacancy statistics $\mathbb{E}[N_b]$ and $\mathbb{E}[n_\emptyset]$. Comparison with Fig.~\ref{fig1results} allows us to characterise this concentration as that where the occupancy first starts to saturate beyond the dilute limit of $0$. We can characterise where this occurs by finding the second solution to $\partial^3_{\ln C_0}\mathbb{E}[N_b]=0$ marking the relevant stabilisation into an occupancy of $\mathbb{E}[N_b]=1$ and vacancy of $\mathbb{E}[n_\emptyset]=0$, on a logarithmic scale. This occurs at characteristic concentration
\begin{align}
\frac{C_0^{\rm (i)}}{C_{\rm eff}}&=\frac{(2+\sqrt{3})K_1K_2}{C^2_{\rm eff}}+\mathcal{O}(\varepsilon^3)
\end{align}
marking the onset of a regime of stable timescales. The next change in regime occurs when the timescales of the system begin to markedly decrease. This can be explained by the bulk concentration being high enough to induce competitive exchange. We can characterise this concentration by considering the probability of dissociation through competitive exchange
\begin{align}
P_{\rm comp}&=\frac{C_0(K_1+k_rK_2)}{K_1(C_0+K_2)+k_rK_2(C_0+K_1)},
\end{align}
and then computing where the rate of increase is changing most dramatically through $\partial^3_{\ln C_0}P_{\rm comp}=0$ giving a second characteristic concentration
\begin{align}
\frac{C_0^{\rm (ii)}}{C_{\rm eff}}&=\frac{(2-\sqrt{3})(1+k_r)K_1K_2}{C_{\rm eff}(K_1+k_rK_2)}+\mathcal{O}(\varepsilon^2).
\label{c12}
\end{align}
Lastly, there is a period of increasing timescales and affinity before each quantity saturates at a high enough concentration. These can all be explained through a saturation of the receptor sites, which we can associate with the largest solution to $\partial^3_{\ln C_0}\mathbb{E}[N_b]=0$
\begin{align}
\frac{C_0^{\rm (iii)}}{C_{\rm eff}}&=(2+\sqrt{3})+\mathcal{O}(\varepsilon).
\end{align}
It is, however, worth dwelling on how the dissociation timescale saturates in this case. To do so we must consider how the mechanisms of competitive exchange are sensitive to concentration. In particular we may appreciate that the timescale of a bound molecule is dependent largely on its ability to perform a given number of avidic `hops' formed of reassociations after one of its binding sites periodically disassociates. The competitor mechanisms increase the dissociation rate by interrupting this process, reducing the number of cyclic hops. As the probability of such an interruption through invasive binding increases, the expected number of hops decreases and the timescale decreases. Consequently when the invasion rate is so high that the expected number of avidic cycles falls to below one, the timescale cannot decrease further and the system saturates. This can be approximated by examining all instances where a singly bound molecule can either fully associate or be blocked by an invading competitor. If we identify some probability $P_{\rm rebind}$ of reassociating from this state the expected number of reassociations before an invasion is given by $\sum_{i=1}^\infty i P^i_{\rm rebind}=P_{\rm rebind}/(1-P_{\rm rebind})^2$. The probability $P_{\rm rebind}$ can be identified by the conditional probability of transitioning to state $D$ or $E$ given state $B$ or $C$. These have probabilities $P_{\rm bind}=C_{\rm eff}/(C_{\rm eff}+C_0+K_{1})$ and $P_{\rm bind}=C_{\rm eff}/(C_{\rm eff}+C_0+K_{2})$ respectively. Assuming $K_2\sim K_1\ll C_{\rm eff}$ we can thus identify $C_0^{\rm (iii)}$ with an expected $9-5\sqrt{3}\sim 0.33$ avidic hops, indicative that the main mechanism by which bivalent molecules increase their stability has been eliminated at such high concentrations.
\par
Next we turn our attention to the $n_1=2$, $n_2=1$ case, illustrated in the panel {\bf d} of SI Fig.~\ref{n11n21}, which exhibits a more complicated set of behaviour, which we describe as five regimes separated by four characteristic concentrations. Similarly to before we observe an initial slowing of association rate at low concentration and a final saturation of all quantities at very high concentration. Characterising these concentrations as the solutions to $\partial^3_{\ln C_0}\mathbb{E}[N_b]=0$ at the first and final stabilisation points of $\mathbb{E}[N_b]$ (c.f. Fig.~\ref{fig1results}) gives similar relations
\begin{align}
\frac{C_0^{\rm (i)}}{C_{\rm eff}}&=\frac{(2+\sqrt{3})K_1K_2}{2C^2_{\rm eff}}+\mathcal{O}(\varepsilon^3)\\
\label{c21}
\frac{C_0^{\rm (iv)}}{C_{\rm eff}}&=2(2+\sqrt{3})+\mathcal{O}(\varepsilon).
\end{align}
What differs here, however, is that $C_0^{\rm (i)}$ also characterises where the timescale of dissociation starts to decrease, precipitating an earlier regime where all quantities are increasing with concentration. We can understand this as being the concentration where occupation is high enough for multi-site exchange to first become as a mechanism of dissociation. The probabilities of multi-site, and regular competitive dissociation in this system are described by 
\begin{widetext}
\begin{align}
P_{\rm ms{-}comp}&=\frac{2 {C_0} {C_{\rm eff}} {K_1}}{2 {K_1} ({C_0} ({C_0}+{C_{\rm eff}}+{K_1})+{K_2} ({C_0}+{K_1}))+{K_2}
   {k_r} ({C_0}+{K_1})^2}\\
   P_{\rm comp}&=\frac{{C_0} (2 {K_1} ({C_0}+{K_1})+{C_0} {K_2} {k_r})}{2 {K_1} ({C_0}
   ({C_0}+{C_{\rm eff}}+{K_1})+{K_2} ({C_0}+{K_1}))+{K_2} {k_r} ({C_0}+{K_1})^2},
\end{align}
\end{widetext}
respectively. As such, we observe that at approximately the same concentration as $C_0^{\rm (i)}$ the probability of multi-site exchange is changing most rapidly around $C_0/C_{\rm eff}=(2+k_r)K_1K_2/2C^2_{\rm eff}+\mathcal{O}(\varepsilon)$.
\par
There are then two further characteristic concentrations defining an intermediate stable regime before a short-lived concentration dependent regime. The latter can be defined as the concentration where conventional competitive exchange shows a significant increase, requiring concentrations around $C_0\sim C_{\rm eff}$ to out compete locally captured competitors engaged in multi-site exchange. We can make this precise by finding the smallest solution to $\partial^3_{\ln C_0}P_{\rm comp}=0$ of order $C_0\sim C_{\rm eff}$ which gives
\begin{align}
\frac{C_0^{\rm (iii)}}{C_{\rm eff}}&=\frac{(4-2\sqrt{3})K_1}{2K_1+k_rK_2}+\mathcal{O}(\varepsilon).
\end{align}
Finally we describe the onset of the intermediate stable regime. This can be understood as a saturation of the multi-site exchange system. In this case saturation is not occurring due to a lower limit on the number of avidic hops, but rather on the rate of competitive invasions that the system can support. Specifically, multi-site exchange relies on a neighbouring receptor site being singly occupied following a periodic partial unbinding by a fully bound molecule. Increases in the probability of this occurring directly leads to larger invasion rates. However, there are a finite number of such neighbouring sites (one in this instance) and so the rate of such invasive binding has an upper limit. In this system the probability is the conditional probability of a primary site being singly occupied given another is singly occupied. This is merely the probability $P_{\rm n}=\langle G|\mathbf{p}^{\rm st}\rangle/(\langle D|\mathbf{p}^{\rm st}\rangle+\langle G|\mathbf{p}^{\rm st}\rangle)=C_0/(C_0+K_1)$. We again characterise the saturation of this probability as the largest solution to $\partial^3_{\ln C_0}P_{\rm n}=0$ yielding
\begin{align}
\frac{C_0^{\rm (ii)}}{C_{\rm eff}}&=\frac{(2+\sqrt{3})K_1}{C_{\rm eff}}+\mathcal{O}(\varepsilon^2).
\end{align}
\section{Kinetics of the $n_1>1$, $n_2=1$ and $n_1=n_2=2$ bipartite configurations}
\label{a4}
Here we utilise Eqs.~(\ref{ea1})-(\ref{ean}) and the definitions from Sec. \ref{a0} to find expressions for the kinetic properties of two, more complicated, bipartite (all-to-all) systems of receptor sites. In particular, the two planar-embeddable configurations $n_1>1$, $n_2=1$ and $n_1=n_2=2$. The first instance leads to the following kinetic properties
\begin{widetext}
\begin{align}
(n_1{+}1)\Gamma_{\rm on}&=\frac{{C_0} {k_1^{\rm on}} \left({K_1} {n_1} \left({C_0^2}+{C_0} ({C_{\rm eff}} ({n_1}-1)+{K_1}+{K_2})+{K_1} {K_2}\right)+{K_2} {k_r} ({C_0}+{K_1})^2\right)}{({C_0}+{K_1})
   ({C_0} {C_{\rm eff}} {n_1}+({C_0}+{K_1}) ({C_0}+{K_2}))}\\
   K_{\rm eff}&=\frac{{K_1} {n_1} ({C_0} ({C_0}-{C_{\rm eff}}+{K_1})+{K_2}
   ({C_0}+{K_1}))+{C_0} {C_{\rm eff}} {K_1} {n_1^2}+{K_2}
   ({C_0}+{K_1})^2}{{n_1} \left({C_0}^2+{C_0} ({K_1}+{K_2})+{K_1}
   ({C_{\rm eff}}+{K_2})\right)+{C_0} {C_{\rm eff}} {n_1^2}+({C_0}+{K_1})^2}\\
   \Gamma_{\rm off}&=\frac{{k^{\rm on}_1} \left({K_1} {n_1} ({C_0} ({C_0}-{C_{\rm eff}}+{K_1})+{K_2} ({C_0}+{K_1}))+{C_0} {C_{\rm eff}} {K_1} {n_1^2}+{K_2} {k_r} ({C_0}+{K_1})^2\right)}{{n_1}
   \left({C_0^2}+{C_0} ({K_1}+{K_2})+{K_1} ({C_{\rm eff}}+{K_2})\right)+{C_0} {C_{\rm eff}} {n_1^2}+({C_0}+{K_1})^2}.
\end{align}
In contrast, for the $n_1=n_2=2$ case we find
\begin{align}
4\Gamma_{\rm on}&=\frac{2 {C_0} {k_1^{\rm on}} \left({C_0^2}{+}{C_0} (2 {C_{\rm eff}}{+}{K_1}{+}{K_2}){+}{K_1} {K_2}\right) ({C_0} ({K_1}{+}{K_2} {k_r}){+}{K_1} {K_2} ({k_r}{+}1))}{{C_0^2} \left(2 {C_{\rm eff}^2}{+}4
   {C_{\rm eff}} ({K_1}{+}{K_2}){+}{K_1^2}{+}4 {K_1} {K_2}{+}{K_2^2}\right){+}(2 {C_0^3}{+}2 {C_0} {K_1} {K_2}) (2 {C_{\rm eff}}{+}{K_1}{+}{K_2})+{C_0^4}{+}{K_1^2}
   {K_2^2}}\\
   K_{\rm eff}&=\frac{({C_0} ({K_1}{+}{K_2}){+}2 {K_1} {K_2}) ({C_0} ({C_0}{+}2
   {C_{\rm eff}}{+}{K_1}){+}{K_2} ({C_0}{+}{K_1}))}{3 {C_0^2} (2
   {C_{\rm eff}}{+}{K_1}{+}{K_2}){+}2 {C_0^3}{+}{C_0} \left(2 {C_{\rm eff}^2}{+}4 {C_{\rm eff}}
   ({K_1}{+}{K_2}){+}{K_1^2}{+}4 {K_1} {K_2}{+}{K_2^2}\right){+}{K_1} {K_2} (2
   {C_{\rm eff}}{+}{K_1}{+}{K_2})}\\
   \Gamma_{\rm off}&=\frac{{k_1^{\rm on}} \left({C_0^2}{+}{C_0} (2 {C_{\rm eff}}{+}{K_1}{+}{K_2}){+}{K_1} {K_2}\right) ({C_0} ({K_1}{+}{K_2} {k_r}){+}{K_1} {K_2} ({k_r}{+}1))}{3 {C_0^2} (2{C_{\rm eff}}{+}{K_1}{+}{K_2}){+}2 {C_0^3}{+}{C_0} \left(2 {C_{\rm eff}^2}{+}4 {C_{\rm eff}} ({K_1}{+}{K_2}){+}{K_1^2}{+}4 {K_1} {K_2}{+}{K_2^2}\right){+}{K_1} {K_2} (2 {C_{\rm eff}}{+}{K_1}{+}{K_2})}.
\end{align}
The behaviour of these systems are illustrated in SI Fig.~\ref{closedfull}. The $n_1>1$ and $n_2=1$ system qualitatively following the behaviour of the $n_1=2$, $n_2=1$ system, but with earlier response with larger numbers of receptor sites. The $n_1=n_2=2$ systems possesses the same number of stable and responsive regimes, but is qualitatively distinct, with the intermediate stable regime being both relatively slow and occurring at lower concentrations compared to the $n_1>1$, $n_2=1$ case. 
\begin{figure*}[!htp]
\centering
\includegraphics[width=0.975\textwidth]{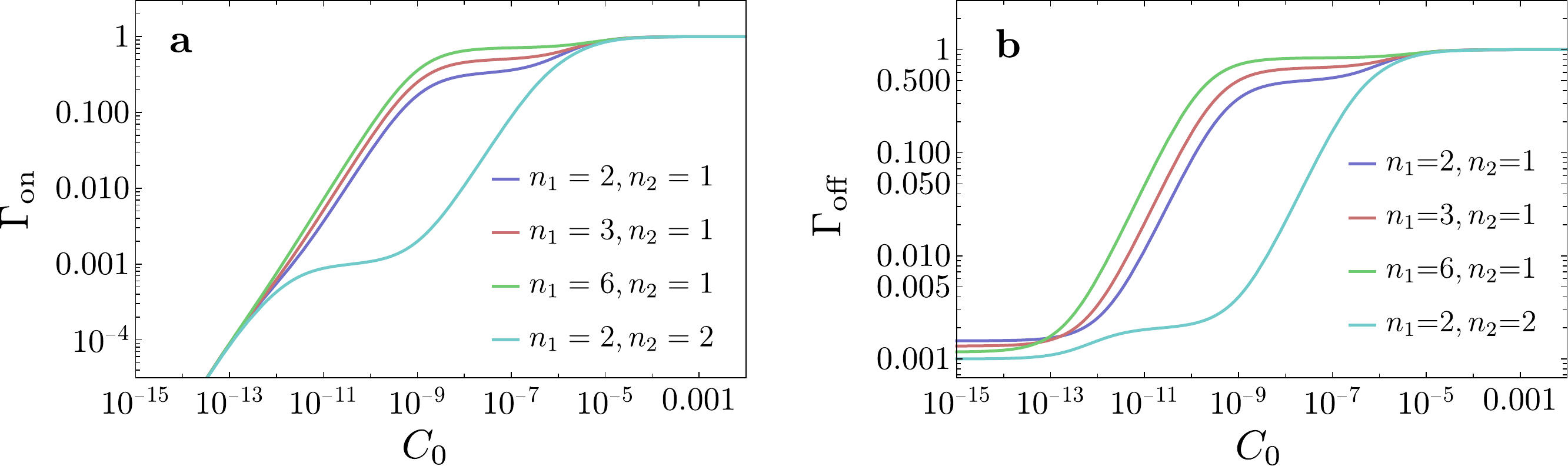}
\caption{Timescales of association (panel {\bf a}) and dissociation (panel {\bf a}) for the $n_1=2,3,6$ and $n_2=1$, and $n_1=n_2=2$ systems. Parameters used are $K_2=K_1=10^{-9}\text{M}$, $C_{\rm eff}=10^{-6}\text{M}$, $k_1^{\rm on}=k_2^{\rm on}=10^9 \text{M}^{-1}s^{-1}$. For these parameters $K_{\rm eff}=\Gamma_{\rm off}/k_1^{\rm on}$.\label{closedfull}}
\end{figure*}
Without disambiguating the mechanisms of dissociation we may still characterise the locations of characteristic concentrations for these systems by solving for maximal log-scale curvature in the dissociation rates $\partial^3_{\ln C_0}\ln \Gamma_{\rm off}=0$. We note that defining the concentrations in this manner leads to different quantitative characterisations than in Sec. \ref{a3} which took a mechanistic approach, however the behaviour with various parameters (e.g. $n_1$ and $K_1$) is unchanged and should be interpreted as the main motivation for deriving such quantities. We may do so with the help of various Ansatzs of the form $C_0=\alpha \varepsilon^m$ and solve to leading order. For $n_1>1$ and $n_2=1$ the first two characteristic concentrations corresponding to the initial rise in both dissociation rate and binding affinity read
\begin{align}
\frac{C_0^{\rm (i)}}{C_{\rm eff}}&=\frac{K_1K_2(k_r+n_1)}{C^2_{\rm eff}(n_1-1)n_1}+\mathcal{O}(\varepsilon^3)\\
\label{cn1n21}
\frac{C_0^{\rm (ii)}}{C_{\rm eff}}&=\frac{K_1}{n_1C_{\rm eff}}+\mathcal{O}(\varepsilon^2).
\end{align}
We note that $C_0^{\rm (i)}$ is also, to leading order, the concentration for which $\Gamma_{\rm off}=2\lim_{C_0\to 0}\Gamma_{\rm off}$. The final two solutions are too unwieldy to report in the general case, but for $K_2=K_1$ and $k_r=1$ read
\begin{align}
\frac{C_0^{\rm (iii)}}{C_{\rm eff}}&=\frac{{n_1} \left(\sqrt{36 {n_1^2}{-}36 {n_1}{+}1}{-}\sqrt{24 {n_1^2}{+}4 \left(\sqrt{36 {n_1^2}{-}36 {n_1}{+}1}{-}6\right) {n_1}{-}2 \sqrt{36 {n_1^2}{-}36 {n_1}{+}1}{+}2}{+}2 {n_1}{-}1\right)}{4 ({n_1}{+}1)}+\mathcal{O}(\varepsilon)\\
\frac{C_0^{\rm (iv)}}{C_{\rm eff}}&=\frac{{n_1} \left(\sqrt{36 {n_1^2}{-}36 {n_1}{+}1}{+}\sqrt{24 {n_1^2}{+}4 \left(\sqrt{36 {n_1^2}{-}36 {n_1}{+}1}{-}6\right) {n_1}{-}2 \sqrt{36 {n_1^2}{-}36 {n_1}{+}1}{+}2}{+}2 {n_1}{-}1\right)}{4 ({n_1}{+}1)}+\mathcal{O}(\varepsilon)
   \end{align}
   corresponding to the end of the second stable regime and the point where the system  entirely saturates at its maximal timescale. These can be reasonably approximated by $C_0^{\rm (iii)}\simeq (n_1-1)C_{\rm eff}/4$ and $C_0^{\rm (iv)}\simeq 7(n_1-1)C_{\rm eff}/2$, respectively. Substituting the geometric mean of the boundaries of the second stable regime ($C_0\to\sqrt{(n_1-1)K_1C_{\rm eff}/4n_1}$) allows us to express the timescale of the intermediate regime, as a fraction of the high $C_0$ limit ($\Gamma_{\rm off}\to (k_2^{\rm off}+n_1 k_1^{\rm off})/(1+n_1)$) as
   \begin{align}
   \frac{\Gamma_{\rm off}^{\rm intermediate}}{\Gamma_{\rm off}^{C_0\to\infty}}&=\frac{n_1^2-1}{n_1(K_2k_r/K_1+n_1)}+\mathcal{O}(\varepsilon^\frac{1}{2}).
   \end{align}
   This ratio approaches $1$ as $n_1$ increases indicating a progressive loss of separating timescales with increasing $n_1$.
   \par
   Similarly, for the $n_1=n_2=2$ case we find
   \begin{align}
   \frac{C_0^{\rm (i)}}{C_{\rm eff}}&=\frac{\left(3+\sqrt{73}-\sqrt{50+6 \sqrt{73}}\right) K_1K_2}{8 C^2_{\rm eff}}+\mathcal{O}(\varepsilon^3)\\
   &\simeq0.1851\frac{K_1K_2}{C^2_{\rm eff}}+\mathcal{O}(\varepsilon^3)\\
      \frac{C_0^{\rm (ii)}}{C_{\rm eff}}&=\frac{\left(3+\sqrt{73}+\sqrt{50+6 \sqrt{73}}\right) K_1K_2}{8 C^2_{\rm eff}}+\mathcal{O}(\varepsilon^3)\\
         &\simeq2.7009\frac{K_1K_2}{C^2_{\rm eff}}+\mathcal{O}(\varepsilon^3)\\
    \frac{C_0^{\rm (iii)}}{C_{\rm eff}}&=\frac{ K_1K_2(1+k_r)}{C_{\rm eff}(K_1+k_rK_2)}+\mathcal{O}(\varepsilon^2)\\
     \frac{C_0^{\rm (iv)}}{C_{\rm eff}}&\simeq 0.354 +\mathcal{O}(\varepsilon)
     \end{align}
     where the final approximate expression arises from the sole positive real root of a $7$th order polynomial. The main distinction here being that the majority of the response occurs between the $C_0^{\rm (iii)}$ and $C_0^{\rm (iv)}$ characteristic concentrations, as opposed to $C_0^{\rm (i)}$ and $C_0^{\rm (ii)}$, for the many to one case.
     \par
     The dilute limit rate of dissociations here is
     \begin{align}
     \Gamma_{\rm off}^{C_0\to 0}&=\frac{k_1^{\rm on}K_1K_2(1+k_r)}{2C_{\rm eff}+K_1+K_2}.
     \end{align}
     We may find the ratio of the intermediate timescale to this quantity by inserting the geometric mean of its boundary concentrations, $C_0^{\rm (ii)}$ and $C_0^{\rm (iii)}$, to find, in this case,
        \begin{align}
   \frac{\Gamma_{\rm off}^{\rm intermediate}}{\Gamma_{\rm off}^{C_0\to 0}}&=2+\mathcal{O}(\varepsilon^{\frac{1}{2}})
   \end{align}
   indicating that there is a stable level of control over the doubling of dissociation rates available at very low concentrations ($C_0\sim K_1K_2/C_{\rm eff}$) through this system. 
   \par
   We have limited ourselves here to the choices of $n_1$ and $n_2$ which are straight-forwardly embeddable when the binding distance is constant, however a relaxation of this property may allow more general systems of receptor sites to be realised. In particular, considering the configuration $n_1=n_2$ reveals a ratio of intermediate to slow timescales of $n_1+\mathcal{O}(\varepsilon^\frac{1}{2})$.
\end{widetext}
\section{One-dimensional chain of receptor sites}
\label{a5}
Here we compute the stationary distribution and kinetic properties of a $1$D chain of receptor sites which bind to a bivalent molecule with identical binding sites, alongside other properties such as its resultant correlation length. To do so we must consider all possible configurations on a line of indistinguishable, equivalent, primary and secondary receptor sites with periodic boundary conditions. The prescription of equivalent receptor sites (Sec.~\ref{a1}) is necessary for the required translational symmetry when using chains formed of odd numbers of receptor sites (it is impossible to have an alternating sequence of two species on a ring for an odd number of elements). This restriction is not required when the ring is formed of an even number of sites, but the method used here does still require the lesser requirement of identical kinetics between the species in that case so that a single transfer matrix, over a unit lattice distance, can be used.
\par
Proceeding in the usual manner by performing a sum in one direction along the lattice sites and one of their neighbours reveals a need to disambiguate not only between whether a molecule bound at a receptor sites is singly or doubly bound, but also the direction along the chain of sites any cross binding is implicated in. This then allows us to assign a probability of zero to combinations which are incompatible. To do this requires $4$ states for each receptor site at location $i$, $r_i\in\{U,B,C_{\leftarrow},C_{\rightarrow}\}$ corresponding to vacancy, singly bound, cross bound to previous site, and cross bound to the next site, respectively. Considering a periodic chain of length $n$, the Boltzmann weighting of some chain configuration $\mathbf{r}\equiv\{r_1,\ldots,r_n\}$ can then be written
\begin{align}
P[\mathbf{r}]&\propto\langle r_1|\mathbf{T}|r_2\rangle\langle r_2|\mathbf{T}|r_3\rangle\ldots\langle r_n|\mathbf{T}|r_1\rangle
\end{align}
which can be achieved with transfer matrix
\begin{align}
\mathbf{T}&=\begin{bmatrix}
\langle U|\mathbf{T}|U\rangle&\langle U|\mathbf{T}|B\rangle&\langle U|\mathbf{T}|C_{\leftarrow}\rangle&\langle U|\mathbf{T}|C_{\rightarrow}\rangle\\
\langle B|\mathbf{T}|U\rangle&\langle B|\mathbf{T}|B\rangle&\langle B|\mathbf{T}|C_{\leftarrow}\rangle&\langle B|\mathbf{T}|C_{\rightarrow}\rangle\\
\langle C_{\leftarrow}|\mathbf{T}|U\rangle&\langle C_{\leftarrow}|\mathbf{T}|B\rangle&\langle C_{\leftarrow}|\mathbf{T}|C_{\leftarrow}\rangle&\langle C_{\leftarrow}|\mathbf{T}|C_{\rightarrow}\rangle\\
\langle C_{\rightarrow}|\mathbf{T}|U\rangle&\langle C_{\rightarrow}|\mathbf{T}|B\rangle&\langle C_{\rightarrow}|\mathbf{T}|C_{\leftarrow}\rangle&\langle C_{\rightarrow}|\mathbf{T}|C_{\rightarrow}\rangle
\end{bmatrix}\nonumber\\
&=\begin{bmatrix}
1 & e^{\Delta F_1/2} & 0 & 1\\
e^{\Delta F_1/2} & e^{\Delta F_1} & 0 & e^{\Delta F_1/2}\\
1 & e^{\Delta F_1/2} & 0 & 1\\
0& 0 &  e^{\Delta F_{c}} & 0
\end{bmatrix}
\end{align}
where  $\Delta F_1=\Delta F_2=\ln C_0/K_1$ is the free energy for a singly bound molecule and $\Delta F_{c}=\ln C_0C_{\rm eff}/K_1^2$ for a cross bound molecule (c.f. Eq.~(\ref{f1})). Zeros correspond to incompatible combinations (e.g. $r_i=U$ and $r_{i+1}=C_{\leftarrow}$, since the latter would require $r_i=C_{\rightarrow}$ to be consistent). 
Summing over all possible combinations yields the trace of $\mathbf{T}^n$. In the limit $n\to \infty$ this allows the free energy \emph{per site} to be expressed as $f=-\beta \ln \lambda_{\rm max}$ where $\lambda_{\rm max}$ is the dominant eigenvalue of $\mathbf{T}$. This is solvable here and yields (again in units $\beta=1$)
\begin{align}
f&=-\ln \lambda_{\rm max}\nonumber\\
&=-\ln\left[\frac{1}{2}\left(1+e^{\Delta F_1}+\sqrt{({1+2e^{\Delta F_1})^2+4e^{\Delta F_{c}}}}\right)\right].
\end{align}
Further, the stationary measure over a single site $|\mathbf{p}^{\rm st}\rangle=\sum_{x\in\{U,B,C_{\leftarrow},C_{\rightarrow}\}}P(x)|x\rangle$ (with $P(x)=\langle x|\mathbf{p}^{\rm st}\rangle$ shorthand for $P(r_i=x)$, valid through translational invariance) can be found using the transfer matrix as follows. By first writing the matrix in terms of its eigendecomposition
\begin{align}
\mathbf{T}=\boldsymbol{\psi}\boldsymbol{\lambda}\boldsymbol{\psi}^{-1}
\end{align}
where $\boldsymbol{\psi}$ is a matrix formed of column vectors equal to the right eigenvectors of $\mathbf{T}$ and $\boldsymbol{\lambda}$ a diagonal matrix with elements equal to its eigenvalues in the order that matches the order of eigenvectors in $\boldsymbol{\psi}$. Consequently, $P(x)$ can be written as the construction
\begin{align}
P(x)&=\frac{\sum_{r_0}\delta _{r_0,x}\langle r_0|\mathbf{T}^n|r_0\rangle}{\sum_{r_0}\langle r_0|\mathbf{T}^n|r_0\rangle}=\frac{\langle x|\boldsymbol{\psi}\boldsymbol{\lambda}^n\boldsymbol{\psi}^{-1}|x\rangle}{\sum_x\langle x|\boldsymbol{\psi}\boldsymbol{\lambda}^n\boldsymbol{\psi}^{-1}|x\rangle}.
\end{align}
This allows us to identify $P(C_{\leftarrow})=P(C_{\rightarrow})=P(C)/2$ with
\begin{widetext}
\begin{align}
P(C)&=\frac{({C_0}{+}{K_1}{+}\eta) ({C_0}{+}{K_1}{-}\eta)^n{-}(C_0{+}K_1{-}\eta) ({C_0}{+}{K_1}{+}\eta)^n}{\eta \left(({C_0}{+}{K_1}{-}\eta)^n{+}({C_0}{+}{K_1}{+}\eta)^n\right)}\\
\eta&=\sqrt{(C_0+K_1)^2+4C_0C_{\rm eff}}.
\end{align}
\end{widetext}
with the probability of other states given simply as $P(U)=K_1(1-P(C))/(C_0+K_1)$ and $P(B)=C_0(1-P(C))/(C_0+K_1)$. The case of the infinite $1$D lattice, $n\to \infty$, can be achieved by taking the appropriate limit or, alternatively, picking out the dominant eigenvalue viz.
\begin{align}
P(x)&={\langle x|\boldsymbol{\psi}|m\rangle\langle m|\boldsymbol{\psi}^{-1}|x\rangle}\nonumber\\
&=\frac{\langle x|\psi^R_{\rm max}\rangle\langle \psi_{\rm max}^L|x\rangle}{\langle \psi_{\rm max}^L|\psi_{\rm max}^R \rangle},
\end{align}
where $m\in\{1,2,3,4\}$ is the index of the column vector that is associated with the dominant eigenvalue. The final expression is expressed in terms of the left and right eigenvectors associated with the dominant eigenvalue of $\mathbf{T}$, $|\psi_{\rm max}^L\rangle$ and  $|\psi_{\rm max}^R\rangle$. We note that it is essential to distinguish between the left and right eigenvectors in this construction, in contrast to usual treatments of, for example, the Ising model where the equivalent expression is of the form $P_x=\langle\psi_{\rm max}^R|x\rangle^2$, due to the asymmetric transfer matrix which in turn is due to the asymmetric coupling resulting from cross-binding. In other words, summing along the chain in the positive direction requires a different matrix (i.e. the transpose) than if summing in the opposite, negative, direction.  Alternatively, one can see this emerging from the coupling on the lattice which, whilst reciprocal, is strictly uni-directional (there is an exclusive interaction with one of its neighbours at any given time).
\par
The relevant eigenvectors for the above transfer matrix are readily found, up to a multiplying constant, as
\begin{align}
|\psi^R_{\rm max}\rangle&=\lambda_{\rm max}e^{-\Delta F_{c}}|U\rangle+\lambda_{\rm max}e^{\Delta F_1/2-\Delta F_{c}}|B\rangle\nonumber\\
&\quad+\lambda_{\rm max}e^{-\Delta F_{c}}|C_{\leftarrow}\rangle+|C_{\rightarrow}\rangle\nonumber\\
|\psi^L_{\rm max}\rangle&=|U\rangle+e^{\Delta F_1/2}|B\rangle+\lambda_{\rm max}^{-1}e^{\Delta F_{c}}|C_{\leftarrow}\rangle+|C_{\rightarrow}\rangle
\end{align}
such that
\begin{align}
|\mathbf{p}^{\rm st}\rangle &\propto \lambda_{\rm max}e^{-\Delta F_{c}}(|U\rangle+e^{\Delta F_1}|B\rangle)+|C_{\leftarrow}\rangle+|C_{\rightarrow}\rangle
\end{align}
with both cross bound states safely symmetric. The probability of finding a receptor site in a cross bound state in this limit is equal to the probability of being in either the $|C_{\leftarrow}\rangle $ or $|C_{\rightarrow}\rangle$ state (i.e. their sum) and is given by
\begin{align}
P(C)&=1-\frac{C_0+K_1}{\sqrt{(C_0+K_1)^2+4C_0C_{\rm eff}}}.
\end{align}
\par
Using such solutions we can compute the aggregate kinetics of the system by recognising that the amortised association rate and mean number of associated molecules are given by
\begin{align}
N\Gamma_{\rm on}&=\mathbb{E}[n^{\emptyset}]C_0k_1^{\rm on}=C_0k_1^{\rm on}nP(U)\\
\mathbb{E}[N_b]&=nP(B)+\frac{nP(C)}{2},
\end{align}
where both can be considered as intensive quantities in the $n\to\infty$  limit by simply dividing through by $n$. We note the factor of two, since there are two individual sites (which the probability concerns) for every cross bound molecule. This then yields kinetic quantities, in terms of $n$,
\begin{align}
   K_{\rm eff}&=\frac{2C_0K_1}{C_0-K_1+\eta\left(1+\frac{2}{(C_0+K_1+\eta)^{n}(C_0+K_1-\eta)^{-n}-1}\right)},\label{n22a}\\
	\Gamma_{\rm on}&=\frac{C_0K_1k_1^{\rm on}}{\eta}\nonumber\\
	&\quad\times\left[1-\frac{2(C_0+K_1-\eta)^n}{(C_0+K_1-\eta)^n+(C_0+K_1+\eta)^n}\right],\\
	\Gamma_{\rm off}&=\frac{2C_0K_1k_1^{\rm on}}{C_0-K_1+\eta\left(1+\frac{2}{(C_0+K_1+\eta)^{n}(C_0+K_1-\eta)^{-n}-1}\right)},\label{n22c}
\end{align}
again with $\eta=\sqrt{(C_0+K_1)^2+4C_0C_{\rm eff}}$. 
We note that for $n=4$ with $K_2=K_1$ these expressions match those in Eqs.~(\ref{n22a})-(\ref{n22c}), as they must, since the $n_1=n_2=2$ all-to-all motif is precisely an $n=4$ chain. Similarly, for $n=3$ the results match those that derive from Eqs.~(\ref{n1equiv}), as the $n=3$ chain is identical to an $n_1=3$ all-to-all motif with equivalent receptor sites.
\par
In the $n\to\infty$ limit these expressions take the simpler forms
\begin{align}
K_{\rm eff}&=\frac{2C_0K_1}{C_0-K_1+\sqrt{4C_0C_{\rm eff}+(C_0+K_1)^2}},\\
\Gamma_{\rm on}&=\frac{C_0K_1k_1^{\rm on}}{\sqrt{4C_0C_{\rm eff}+(C_0+K_1)^2}},\\
\Gamma_{\rm off}&=\frac{2C_0K_1k_1^{\rm on}}{C_0-K_1+\sqrt{4C_0C_{\rm eff}+(C_0+K_1)^2}}.
\end{align}
The response to bulk concentration in this system depends strongly on the value of $n$ as illustrated in Fig.~2 in the main text. For small, odd, $n$ the behaviour qualitatively follows that of the $n_1>1$ $n_2=1$ all-to-all motif with multi-site exchange directly implicated due to the fact that even when double binding is strongly preferred thermodynamically, this system is always `frustrated' due to the existence of a spare receptor site which can be occupied in only a singly bound way. Simply: batons cannot perfectly tile on the chain. For small, even, $n$ the behaviour follows that of the $n_1=n_2=2$ all-to-all motif (indeed they are equal for $n=4$) since molecules \emph{can} tile perfectly in this case. However, as $n$ is increased the two results converge on a distinct behaviour where no intermediate timescale exists, characterised by a wider scale of concentrations giving a much shallower and broader tunable region. In this $n\to \infty$ limit, by solving for maximal curvature on a log scale ($\partial^3_{\ln C_0}\Gamma_{\rm off}=0$), we can identify the boundaries of this tunable region to be given by
\begin{align}
C_0=C_{\rm eff}+2K_1\pm\sqrt{(C_{\rm eff}+2K_1)^2-K_1^2}
\end{align}
which can be asymptotically identified with $C_0/C_{\rm eff}=K_1^2/2C^2_{\rm eff}+\mathcal{O}(\varepsilon^3)$ and $C_0/C_{\rm eff}=2+\mathcal{O}(\varepsilon)$.
\par
In this limit the presence (or absence) of double binding which characterises the $n_1>1$ $n_2=1$ and $n_1=n_2=2$ all-to-all motifs respectively, is no longer determined by the finite size of the chain and the explicit possibility of tiling, but rather on the many body co-ordination along the lattice and the likelihood of a `domain boundary' between perfect tilings of batons which can act as a site for multi-site exchange (Fig.~2{\bf d} in the main text). The likelihood of these multi-site exchange sites occurring is related to the correlation that can persist along the chain, providing us with a first example of an explicit many body property that the avidic binding may give rise to on such extended systems.
\par
Moving forwards, we can determine the correlation length in this system by first finding conditional probabilities of receptor site states at neighbouring sites by constructing the joint probabilities between two sites, separated by $k$ receptor sites, as 
\begin{align}
P(r_i{=}x,r_{i{+}k}{=}y)&=\lim_{N\to\infty}\frac{\langle x|\mathbf{T}^k|y\rangle\langle y|\mathbf{T}^{N-k}|x\rangle}{Z}\nonumber\\
&=\lambda_{\rm max}^{-k}\langle x|\mathbf{T}^k|y\rangle\langle y|\boldsymbol{\psi}|m\rangle\langle m|\boldsymbol{\psi}^{-1}|x\rangle.
\end{align}
For the case $y=x$ this becomes
\begin{align}
P(r_i{=}x,r_{i{+}k}{=}x)&=\lambda_{\rm max}^{-k}\langle x|\mathbf{T}^k|x\rangle\langle x|\boldsymbol{\psi}|m\rangle\langle m|\boldsymbol{\psi}^{-1}|x\rangle\nonumber\\
&=\lambda_{\rm max}^{-k}\langle x|\mathbf{T}^k|x\rangle P(x)
\end{align}
and so we can identify the conditional probability
\begin{align}
P(r_{i{+}k}{=}x|r_{i}{=}x)&=\lambda_{\rm max}^{-k}\langle x|\mathbf{T}^k|x\rangle.
\end{align}
For $x=U$ this gives, for $k>0$,
\begin{align}
&P(r_i{=}U|r_{i{+}k}{=}U)=\nonumber\\
&P(U)\left[1-\left(\frac{C_0+K_1-\sqrt{(C_0+K_1)^2+4C_0C_{\rm eff}}}{C_0+K_1+\sqrt{(C_0+K_1)^2+4C_0C_{\rm eff}}}\right)^k\right].
\end{align}
In turn this allows us to identify the (inverse of the) correlation length of the system, which we define through
\begin{align}
&l_{\rm corr}^{-1}=\nonumber\\
&\lim_{k\to\infty}-\frac{1}{k}\ln|P(r_i{=}U,r_{i{+}k}{=}U)-P(r_i{=}U)P(r_{i{+}k}{=}U)|,
\end{align}
as
\begin{align}
l^{-1}_{\rm corr}&=\ln\left[\frac{\sqrt{(C_0+K_1)^2+4C_0C_{\rm eff}}+C_0+K_1}{\sqrt{(C_0+K_1)^2+4C_0C_{\rm eff}}-C_0-K_1}\right]
\end{align}
in units of number of receptor sites. Its behaviour is shown in panel {\bf b} of Fig.~2 in the main text. 
This vanishes at $C_0\to 0$ and $C_0\to \infty$ where cross binding is absent and each site is independent, and peaks at $C_0=K_1$ where the system most closely achieves a perfect tiling of cross bound molecules along the chain, with maximum correlation length
\begin{align}
l^{\rm max}_{\rm corr}&=\frac{C^{1/2}_{\rm eff}}{2K^{1/2}_1}+\frac{K^{1/2}_{1}}{12C^{1/2}_{\rm eff}}+\mathcal{O}(\varepsilon^{\frac{5}{2}}).
\end{align}
It is not difficult to see that on such a system batons can diffuse along the chain. But the picture is complicated by the fact that the batons will experience both exclusion effects as well as an increased likelihood of returning to bulk whenever they encounter another baton on the chain due to the multi-site exchange phenomenon. We can, however, estimate the mean run length for this phenomenon in the following way. First, we can describe the timescale of a directed step as the time required to unbind a single end of a baton and then rebind to a distinct site, whilst the other end of the baton remains associated with another receptor site. In one dimension this involves one end of the baton unbinding, the baton pivoting through 180 degrees, and then rebinding to the neighbour on the opposite side of the baton which remained bound. We can describe this as the mean time to unbind and rebind to any site multiplied by the expected number of such events that lead to a directed step. We may approximate these quantities by considering the neighbouring sites fixed during such an event. Assuming, without loss of generality, that it is the left most site which has dissociated, resulting in a directed step when it rebinds on the right hand side, the approximation of the mean time to unbind and rebind on either side can be written $t_{\rm rebind}\simeq (k_1^{\rm off})^{-1}+((1+P_{U|C_{\leftarrow}})C_{\rm eff}k_1^{\rm on})^{-1}$ and the expected number of such events before a directed step as $n_{\rm rebind}\simeq P_{U|C_{\leftarrow}}/(1+P_{U|C_{\leftarrow}})$ where $P_{U|C_{\leftarrow}}$ is the probability of the site on the right hand side (in the $\rightarrow$, positive direction) neighbouring the cross bound molecule being vacant at the time of unbinding, equal to $2K_1/(C_0+K_1+\sqrt{(C_0+K_1)^2+4C_0C_{\rm eff}})$, found using the above expressions for conditional probabilities. In both cases the expressions $1+P_{U|C_{\leftarrow}}$ represent the expected number of neighbouring vacant sites - one is assumed to exist with probability one as it has just been vacated, whilst the other uses the conditional probability. It is the latter that results in directed motion. The expected time for a directed step is then simply $t_{\rm rebind}n_{\rm rebind}$. One then can crudely estimate the expected squared displacement, or run-length, by taking the ratio of the expected lifetime ($\Gamma_{\rm off}^{-1}$) to the step time, since the system obeys detailed balance by construction and so is diffusive in the sense of an unbiased 1D random walk. This then gives
\begin{widetext}
\begin{align}
l_{\rm run}^2\sim\frac{K_1C_{\rm eff}\left(C_0-K_1+\sqrt{(C_0+K_1)^2+4C_0C_{\rm eff}}\right)}{C_0\left[2K_1(2C_{\rm eff}+K_1)+(C_{\rm eff}+K_1)\left(C_0-K_1+\sqrt{(C_0+K_1)^2+4C_0C_{\rm eff}}\right)\right]},
\end{align}
\end{widetext}
illustrated in panel {\bf b} of Fig.~2 in the main text.
\section{Nearest neighbour dependence model for lattice configurations}
\label{a6}
\begin{figure*}[!htp]
\centering
\includegraphics[width=0.95\textwidth]{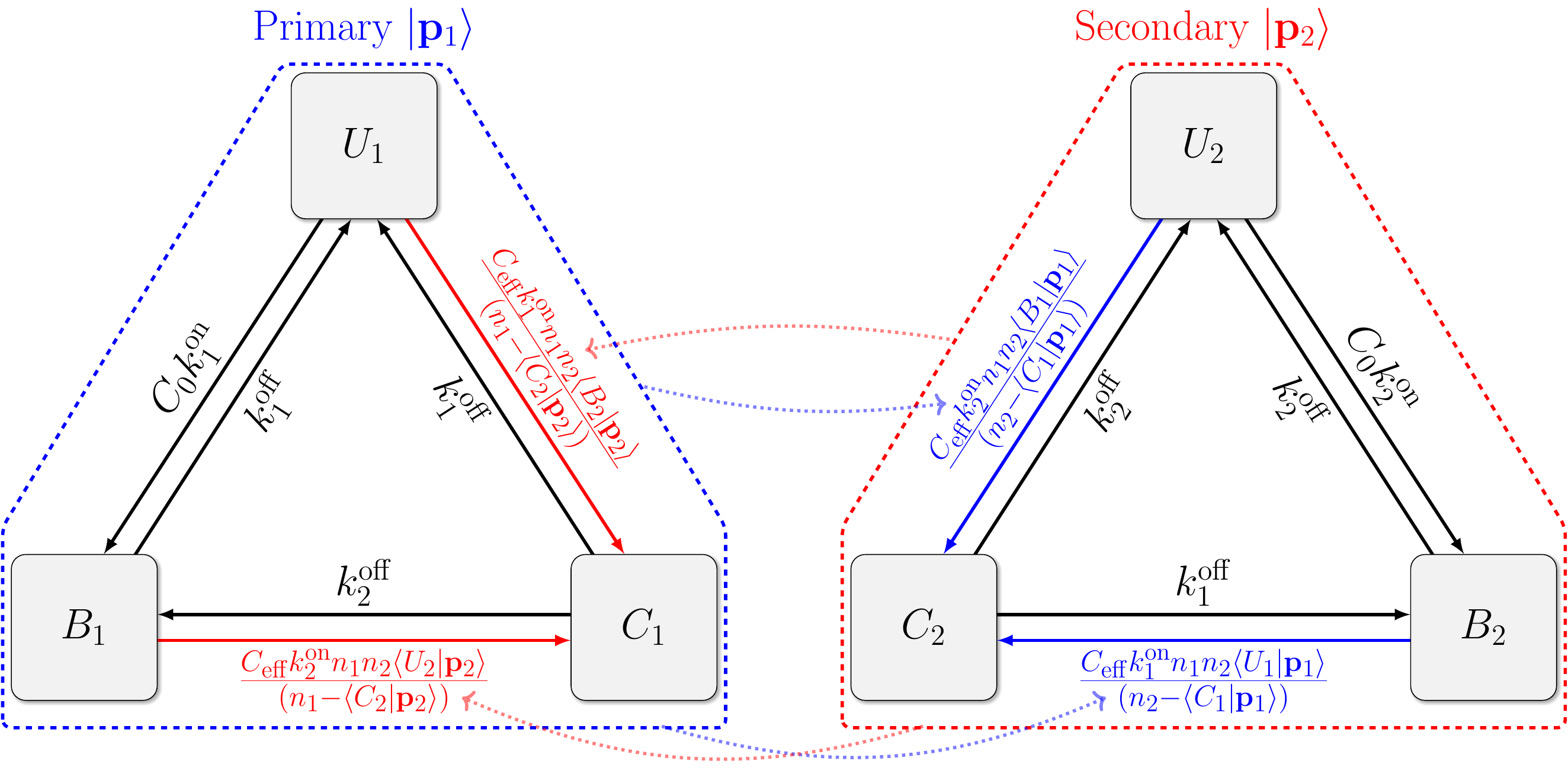}
\caption{State diagram, allowed transitions, and transition rates for the model approximating the kinetics of extended systems where $n_1$ and $n_2$ are taken as co-ordination numbers of interleaved lattices. The model consists of two coupled systems describing the mean occupation statistics of primary and secondary sites respectively. The system contained in the dotted blue region describes the primary sites which are characterised through probability vector $|\mathbf{p}_1\rangle$. Similarly, the secondary sites are characterised by the probability vector $|\mathbf{p}_2\rangle$ over the system contained within the dotted red region. Coupling between the systems is achieved through transition rates associated with cross-binding and are drawn as solid red/blue arrows. Transitions drawn in blue change the state of a secondary receptor site, but depend on the probability vector for the primary system $|\mathbf{p}_1\rangle$, and transitions drawn in red change the state of a primary receptor site, but depend on the probability vector for the secondary system $|\mathbf{p}_2\rangle$, with such dependence emphasised with the dotted blue/red arrows respectively.\label{mf}}
\end{figure*}

Here we construct a model description for infinite  extended configurations of receptor sites consisting of interleaved lattices of primary and secondary sites by considering the effective dynamics of a randomly chosen pair of  primary and secondary sites. Each receptor site can exist in three states relating to its occupation. The primary site can exist in states $U_1$, ${B}_1$, and $C_1$ corresponding to vacancy, occupation by a molecule which is not bound at its opposing binding site, and occupation by a molecule which is also bound at its opposing binding site to any secondary receptor site, respectively. Similarly the secondary site can exist in states $U_2$, $B_2$, and $C_2$, with the role of primary and secondary receptor sites reversed. The dynamics of the primary site are given by a $3\times 3$ generator $\mathbf{T}_1$ operating on state vector $|\mathbf{p}_1\rangle$, and the secondary site by a $3\times 3$ generator $\mathbf{T}_2$ operating on state vector $|\mathbf{p}_2\rangle$, with constituent rates illustrated in SI Fig.~\ref{mf}. The total behaviour is then equally described as two non-autonomous, coupled, linear systems or as a joint, autonomous, and non-linear, system with generator $\mathbf{T}_{12}=\mathbf{T}_1\oplus\mathbf{T}_{2}=\mathbf{T}_1\otimes\mathbb{I}+\mathbb{I}\otimes\mathbf{T}_2$ operating on state vector $|\mathbf{p}_{12}\rangle$ over $\{U_1,B_1,C_1\}\times\{U_2,B_2,C_2\}$. The non-linearity follows crucially from the rates in both $\mathbf{T}_{1}$ and $\mathbf{T}_{2}$ depending on the state vectors, $|\mathbf{p}_2\rangle$ and $|\mathbf{p}_1\rangle$, respectively. 
\par
Whilst transitions $C\to B$, $C\to U$, and $U\leftrightarrow B$ are straight-forward, the specific form of the $B\to C$ and $U\to C$ rates warrant explanation. Let us consider, without loss of generality, the transition $B_1\to C_1$. In order to transition from the singly bound to doubly bound state there must be a vacant secondary receptor site which neighbours the selected primary site, with the rate proportional to their number. Since the rates into different configurations of secondary receptor sites accumulates linearly in proportion to their probability this is simply given by $n_2$ multiplied by the conditional probability of selecting a single unbound neighbouring secondary site, given the knowledge that one primary site is occupied. Naively, in mean field, these events are independent and so the number is given by $n_2\langle U_2|\mathbf{p}_2\rangle$, and thus the rate would be given by $C_{\rm eff}k_2^{\rm on}n_2\langle U_2|\mathbf{p}_2\rangle$. This amounts to assuming that the primary and secondary sites are distantly separated and thus independent, which is of course false since they must be immediate neighbours to facilitate cross-binding. Proceeding with the naive interpretation leads to dynamics which fail dramatically even in the $n_1=n_2=1$ case. Explicitly, it is not capturing correlations between immediate neighbours.
\par
To correct for this correlation we recognise that it manifests only through cross-binding between adjacent receptor sites. This possibility of cross-binding confers some additional information about the state of a randomly chosen neighbouring secondary site  when the state of a primary receptor site is known as it can eliminate combinations relating to doubly bound molecules. For instance in the $n_1=n_2=1$ case there is some probability of a primary site being occupied by a cross bound molecule $\langle C_1|\mathbf{p}_1\rangle$. However, if it is known that a secondary site is vacant, then the conditional probability of the primary site being cross-bound is zero, not $\langle C_1|\mathbf{p}_1\rangle$ since there are no compatible secondary sites for such a cross bound molecule to be bound with. We may denote the marginal solutions from the model $P_i^{\rm M}(x_i)=\langle x_i|\mathbf{p}_{i}\rangle$ and note that as described it possesses independent joint distributions $P_{12}^{\rm M}(x_1,y_2)=\langle x_1|\mathbf{p}_{1}\rangle\langle y_2|\mathbf{p}_{2}\rangle$ corresponding to randomly chosen primary and secondary sites from the entire lattice. However, the joint probabilities for immediate neighbours differs from this expression, $P_{12}(x_1,y_2)\neq P^{\rm M}_{12}(x_1,y_2)$. We can, however, find appropriate corrections to $P^{\rm M}_{12}(x_1,y_2)$ to approximate $P_{12}(x_1,y_2)$ used in the $B\to C$ and $U\to C$ transitions. We formulate such a correction as follows. 
\par
Given a randomly selected pair of neighbouring primary and secondary receptor sites,  we identify the specific receptor sites $s_1\in\mathcal{S}_1=\{1,\ldots,n_1\}$ and $s_2\in\mathcal{S}_2=\{1,\ldots,n_2\}$ as members of the set $\mathcal{S}_1$ of primary receptor sites that neighbour the chosen secondary site and the set $\mathcal{S}_2$ of secondary receptor sites that neighbour the chosen primary site. We can then appreciate that the mean field solutions $\langle x_2|\mathbf{p}_2\rangle$ are equivalent to the joint probabilities $P_2(s_2\in\mathcal{S}_2,x_2)=\sum_{i\in\mathcal{S}_2}P_2(s_2=i|x_2)P_2(x_2)=\sum_{i\in\mathcal{S}_2}P_2(s_2=i)P_2(x_2)=P_2(x_2)$, where $P_2(s_2=i|x_2)=P_2(s_2=i)$ since the receptor sites are assumed to be identical. The probability $P_2(s_2=i)$ is arbitrary, but we will assume a flat prior for all similar inferences such that $P_2(s_2=i)=n_2^{-1}$, consistent with the sites being indistinguishable. Consequently, when, for example, considering the transition rate for $B_1\to C_1$ we require the expected number of vacant sites for the secondary binding site to associate with. This is proportional to the true conditional probability $P_2(s_2\in\mathcal{S}_2,U_2|s_1\in\mathcal{S}_1,B_1)$ describing the probability a randomly selected secondary site is vacant given the knowledge that a \emph{single} neighbouring primary site, $s_1$, is known to be singly bound.  Without accounting for neighbour to neighbour cross-binding, the model incorrectly factors this as $P^{\rm M}_2(s_2\in\mathcal{S}_2,U_2|s_1\in\mathcal{S}_1,B_1)=P_2(s_2\in\mathcal{S}_2,U_2)$. Instead we thus manually construct an expression for the conditional probability $P_2$ in terms of expressions for the joint probabilities $P_{12}$, which in turn are described in terms of the native model solutions $P^{\rm M}_{12}$, using simple combinatorial corrections.
\par
We compute the desired conditional probability in the usual manner, viz.
\begin{align}
&P_2(s_2\in\mathcal{S}_2,U_2|s_1\in\mathcal{S}_1,B_1)\nonumber\\
&=\frac{P_{12}(\{s_2\in\mathcal{S}_2,U_2\},\{s_1\in\mathcal{S}_1,B_1\})}{P_1(s_1\in\mathcal{S}_1,B_1)}\\
&=\frac{P_{12}(\{s_2\in\mathcal{S}_2,U_2\},\{s_1\in\mathcal{S}_1,B_1\})}{\sum_{x\in\{U_2,B_2,C_2\}}P_{12}(\{s_2\in\mathcal{S}_2,x\},\{s_1\in\mathcal{S}_1,B_1\})}.
\end{align}
Let us consider the various terms. First we consider $P_{12}(\{s_2\in\mathcal{S}_2,U_2\},\{s_1\in\mathcal{S}_1,B_1\})$. These binding combinations do not directly interfere with each other, and so the independence of the receptor sites in the model allows this to be written
\begin{align}
&P_{12}(\{s_2\in\mathcal{S}_2,U_2\},\{s_1\in\mathcal{S}_1,B_1\})\nonumber\\
&\propto P^{\rm M}_{12}(\{s_2\in\mathcal{S}_2,U_2\},\{s_1\in\mathcal{S}_1,B_1\})\\
&=\kappa \sum_{i\in\mathcal{S}_2}\sum_{j\in\mathcal{S}_1}P_2(s_2=i|{U}_2)P_2({U}_2)P_1(s_1=j|B_1)P_1(B_1)\\
&=\kappa P_2({U}_2)P_1(B_1)
\end{align}
where $\kappa$ is some proportionality constant and, as before, we take $P(s_1=j|U_2)=P(s_1=j)$. Analogously $P_{12}(\{s_2\in\mathcal{S}_2,B_2\},\{s_1\in\mathcal{S}_1,B_1\})=\kappa P_2(B_2)P_1(B_1)$.
\par
However for the equivalent expression over the state $C_2$ the binding states are \emph{not} independent, and we need to account for the fact that it represents a binding across a specific pair of one secondary \emph{and} primary receptor site, denoted $\{s_2,s'_2\}\in\mathcal{S}_2 \times\mathcal{S}_1$, which rules out combinations where the site $s'_2=s_1\in\mathcal{S}_1$ is not cross-bound. We write this as
\begin{align}
&P_{12}(\{s_2\in\mathcal{S}_2,C_2\},\{s_1\in\mathcal{S}_1,B_1\})\nonumber\\
&\propto(1-\delta_{s_1s'_2})P^{\rm M}_{12}(\{s_2\in\mathcal{S}_2,s'_2\in\mathcal{S}_1,C_2\},\{s_1\in\mathcal{S}_1,B_1\})\\
&=\kappa\sum_{i\in\mathcal{S}_2}\sum_{j\in\mathcal{S}_1}\sum_{k\in\mathcal{S}_1}(1-\delta_{jk})P_2(s_2=i,s'_2=j|C_2)\nonumber\\
&\qquad\times P_1(s_1=k|B_1)P_2(C_2)P_1(B_1)\\
&=\kappa\frac{1}{n^2_1}\sum_{j\in\mathcal{S}_1}\sum_{k\in\mathcal{S}_1}(1-\delta_{jk})P_2(C_2)P_1(B_1)\\
&=\kappa\frac{(n_1-1)P_2(C_2)P_1(B_1)}{n_1},
\end{align}
where we have taken the flat priors $P(s_1=k)=n_1^{-1}$, $P(s_2=i,s'_2=j)=(n_1n_2)^{-1}$.  
Consequently, the conditional probability associated with the transition can be expressed
\begin{align}
P_2(s_2\in\mathcal{S}_2,U_2|s_1\in\mathcal{S}_1,B_1)&=\frac{P_{2}(U_2)}{1-P_2(C_2)/n_1}\nonumber\\
&=\frac{n_1\langle U_2|\mathbf{p}_{2}\rangle}{n_1-\langle C_2|\mathbf{p}_2\rangle}.
\label{truecond}
\end{align}
This in turn identifies the proportionality constant as $\kappa=(1-\langle C_2|\mathbf{p}_2\rangle/n_1)^{-1}$. 
As such the transition rate for $B_1\to C_1$ is given by $C_{\rm eff}k_2^{\rm on}n_1n_2\langle U_2|\mathbf{p}_2\rangle/(n_1-\langle C_2|\mathbf{p}_2\rangle)$ as indicated, 
with analogous corrections for all other such marked transitions. We note that the moderating pre-factor introduced to the rate must lie in $[1,n_1/(n_1-1)]$ meaning that for larger co-ordination numbers the correlation effects become reduced as the interaction is uni-directional (a baton can only be physically be bound to one neighbouring site), so the effect becomes diluted.
\par
Given such dynamics, we then seek the stationary vectors $|\mathbf{p}_1^{\rm st}\rangle$ and $|\mathbf{p}_2^{\rm st}\rangle$, characterising mean occupancies in the steady state.  This can be expedited by noting that consistency requires $n_1\langle C_1|\mathbf{p}_1\rangle=n_2\langle  C_2|\mathbf{p}_2\rangle$, and that an absence of aggregate currents in the stationary state necessitate detailed balance such that
\begin{align}
C_0\langle U_1|\mathbf{p}^{\rm st}_1\rangle&=K_1\langle B_1|\mathbf{p}^{\rm st}_1\rangle\nonumber\\
C_0\langle U_2| \mathbf{p}^{\rm st}_2\rangle&=K_2\langle B_2|\mathbf{p}^{\rm st}_2\rangle.
\end{align}
By introducing the parameters
\begin{align}
\gamma &=\frac{{n_1} {n_2} \left({C_0} (C_{\rm eff} ({n_1}+{n_2})+{K_1}+{K_2})+{C_0}^2+{K_1} {K_2}\right)}{2 C_{\rm eff} {C_0} {n_1} {n_2}+2 ({C_0}+{K_1}) ({C_0}+{K_2})}\\
\beta&=\frac{C_0C_{\rm eff}n_1^2n_2^2}{(C_0+K_1)(C_0+K_2)+C_0C_{\rm eff}n_1n_2}
\end{align}
 we can describe the mean number of cross-bound molecules per $(n_1+n_2)$ receptor sites, $N_{c}$, as
\begin{align}
\mathbb{E}[N_{c}]&=n_1\langle C_1|\mathbf{p}^{\rm st}_1\rangle=n_2\langle C_2|\mathbf{p}^{\rm st}_2\rangle\nonumber\\
&=\gamma-\sqrt{\gamma^2-\beta}
\label{e12}
\end{align}
which can be used to characterise the stationary solutions and number of expected molecules (per $n_1+n_2$ receptor sites) in any given configuration
\begin{align}
\mathbb{E}[n_{1}^B]&=n_1\langle B_1|\mathbf{p}^{\rm st}_1\rangle=\frac{C_0(n_1-\mathbb{E}[N_{c}])}{C_0+K_1}\\
\mathbb{E}[n_{1}^\emptyset]&=n_1\langle U_1|\mathbf{p}^{\rm st}_1\rangle=\frac{K_1(n_1-\mathbb{E}[N_{c}])}{C_0+K_1}\\
\mathbb{E}[n_{2}^B]&=n_2\langle B_2|\mathbf{p}^{\rm st}_2\rangle=\frac{C_0(n_2-\mathbb{E}[N_{c}])}{C_0+K_2}\\
\mathbb{E}[n_{2}^\emptyset]&=n_2\langle U_2|\mathbf{p}^{\rm st}_2\rangle=\frac{K_2(n_2-\mathbb{E}[N_{c}])}{C_0+K_2}.
\label{En}
\end{align}
Consequently, the probability of a receptor being in a singly bound state is merely an equilibrium occupation (i.e. of the form $C_0/(C_0+K_{\cdot})$) once the appropriate number of sites removed from the system by cross-linking, $\mathbb{E}[N_{c}]$, are discounted.
\par
The solutions given in Eqs.~(\ref{e12})-(\ref{En}), and the subsequent conditional probabilities in Eq.~(\ref{truecond}), are \emph{exact} in two of the cases we have seen already, namely the $n\to\infty$ linear chain where $K_2/K_1=k_r=1$ in Sec.~\ref{a5} replicated by setting $n_1=n_2=2$, and the many-to-one motif in Sec.~\ref{a4} where the receptor site number and co-ordination number descriptions, $n_1\ge 1$, $n_2=1$, are equivalent.
\par
The exact results in these two cases can be understood by appreciating that the model is exact when the conditional probabilities utilised in the transitions rates are accurate. Interactions, and thus correlations, that determine these conditional probabilities on such lattices emerge from the statistical influence of doubly bound molecules. As such the conditional probabilities depend on all chains of non-repeating, alternating, primary and secondary receptor sites (between which molecules can be doubly bound) that can be drawn between a given receptor site and its immediate neighbours. In the case $n_1\geq 1$, $n_2=1$ there exist no correlation chains other than the one link case between immediate neighbours calculated above and so the result is exact. Similarly in the $n_1=n_2=2$ infinite chain there exists the one link chain, calculated above, and one other with length equal to the size of the system when periodic boundaries are implemented, which decays to zero as the lattice size is increased, and so the correction again is exact. Both these cases are equivalent to the absence of (finite) loops from any given receptor site giving some additional insight into the workings of the approximation. Clearly, with embeddings into higher dimensions there exists multiple correlation chains (loops) between any neighbouring pairs of receptor sites and the model becomes a first order approximation to all such contributions. However, the model allows us to exactly describe the statistics of the 1D chain with asymmetric receptor sites (something that was not possible with the transfer matrix which assumed equivalent [odd $n$] or symmetric, i.e. $K_2=K_1$ [even $n$ and $n\to\infty$] receptor sites), and in practice provides an excellent approximation for embeddings in higher dimensions for the parameters used.
\par
 Examples of where such lattices are embeddable into a plane are shown in panel {\bf a} of Fig.~3 in the main text, illustrating how $n_1$ is to be interpreted as effective co-ordination number of primary sites that any given secondary sites (drawn as circles) can interact with through cross-binding, whilst $n_2$ is to be interpreted similarly as the effective co-ordination for the primary sites (drawn as squares). Accuracy of the solution is tested against numerics provided by an importance sampling algorithm for the $n_1=n_2=4$ and $n_1=4, n_2=2$ cases in panel {\bf b} of Fig.~3 in the main text where we see almost perfect agreement. Such high accuracy can be attributed to the low weight of the influence of higher order loops when co-ordination numbers are large. A crude estimate of the contribution from a loop, given the correction of the one-step correlation calculated above, is $\sim (n_1n_2)^{-k/2}$ where $k$ is the length of the loop, in receptor sites, from a site to one of its nearest neighbour (for which the conditional probability is being estimated for). The smallest size loop for square, $n_1=n_2=4$, lattices is $k=3$, whilst for the $n_1=4, n_2=2$ case the smallest loop is $k=7$,  leading to estimated small corrections of $\sim 1.6\%$ and $\sim 0.07\%$, respectively, which are the order of the differences we see with numerics. This simple estimate would suggest that accuracy may be lower for triangular and hexagonal embeddings where co-ordination numbers and loops as small as $n=3$ and $k=3$ can exist. 
 \par
 The importance sampling algorithm consists of selecting a random pair of neighbouring primary and secondary receptor sites, proposing a random pair of states in $\{U,B,C_{\uparrow},C_{\downarrow},C_{\leftarrow},C_{\rightarrow}\}^2$ where the arrow designates the direction to the nearest neighbour on the square lattice to which the molecule is also bound, and accepting/rejecting according to the free energy difference through a Metropolis-Hastings criterion. Note, incompatible configurations (those with cross bound state, $C_{\uparrow},C_{\downarrow},C_{\leftarrow},C_{\rightarrow}$, not matched by a suitable neighbour, $C_{\downarrow},C_{\uparrow},C_{\rightarrow},C_{\leftarrow}$) possess infinite free energy and a vanishing probability, allowing the proposal of such configurations to be elided entirely in this scheme. The results in Fig.~3{\bf b} are obtained using a $300\times300$ lattice with periodic boundary conditions. The $n_1=4, n_2=2$ lattice utilises a $300\times 300$ square lattice ($n_1=n_2=4$) with appropriate sites removed  such that $3/4$ of the total sites remain.
\par
\begin{figure}[!htp]
	\centering
	\includegraphics[width=0.48\textwidth]{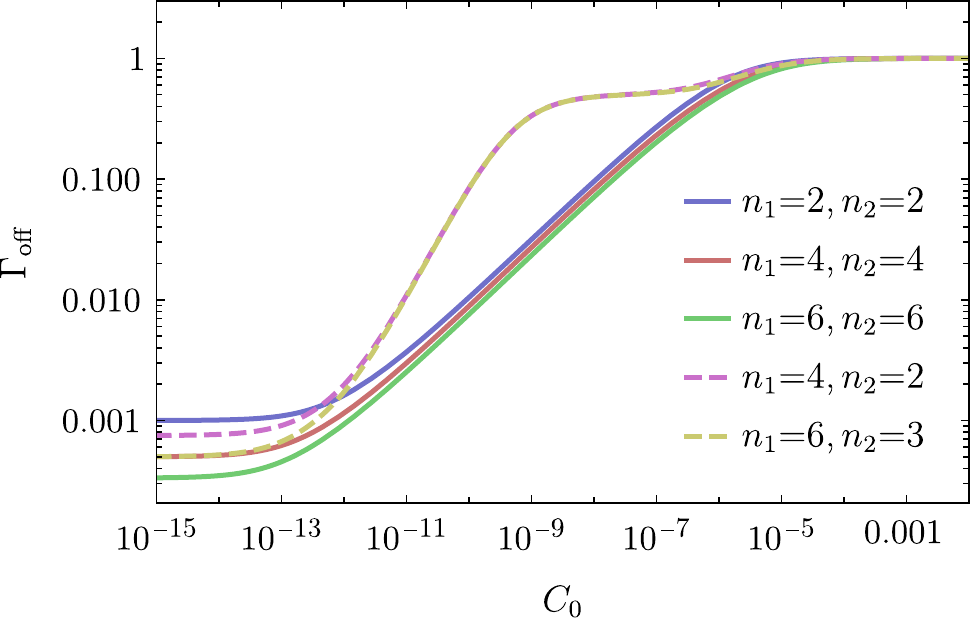}
	\caption{dissociation rates per bound molecule for lattices with different co-ordination numbers. When co-ordination numbers are equal between species (solid lines) the behaviour follows that of the infinite 1$D$ chain with multi-site exchange controlled by the break down of perfect tiling on the lattice. In contrast with different co-ordination numbers (dashed lines), despite the infinite size, the behaviour follows the many to one motif as the unequal lattices cause frustration at every point on the lattice, allowing sites for multi-site exchange. 
Parameters used are $K_2=K_1=10^{-9}\text{M}$, $C_{\rm eff}=10^{-6}\text{M}$, $k_1^{\rm on}=k_2^{\rm on}=10^9 \text{M}^{-1}s^{-1}$.\label{latticerates}
	}
\end{figure}

The kinetic properties are then simply calculated, as before, since the number of primary and secondary sites exists in number proportional to $n_1$ and $n_2$. The resulting behaviours are qualitatively similar to those which we have already observed, with two distinct behaviours for equal and distinct co-ordination numbers, however the change in co-ordination number quantitatively changes some of the key properties of the system. The dissociation rates are illustrated for several lattice configurations in SI Fig.~\ref{latticerates}.
\par
 The most dramatic is the change to the stability under dilution which we can write as
\begin{align}
\Gamma_{\rm off}^{\rm dilute}&=\lim_{C_0\to 0}\Gamma_{\rm off}=\frac{k_1^{\rm on}K_1  K_2 (n_1 + k_r n_2)}{K_2 n_1 +K_1n_2+ C_{\rm eff} n_1 n_2}.
\label{Goffopenn1n2}
\end{align}
with the product of the co-ordination numbers in the denominator allowing for much longer bound lifetimes than observed previously.
\par
In turn, for the $n_1\neq n_2$ we may estimate the ratio of the observed second and third stable timescales through an evaluation of the dissociation rate at $C_0\to \sqrt{C_{\rm eff}K_1}$ yielding
\begin{align}
   \frac{\Gamma_{\rm off}^{\rm intermediate}}{\Gamma_{\rm off}^{C_0\to \infty}}&=\frac{n_1^2-n_2^2}{n_1^2+\frac{k_2^{\rm off}n_1n_2}{k_1^{\rm off}}}+\mathcal{O}(\varepsilon^{\frac{1}{2}}).
   \end{align}
   This broadly behaves as $(n_1-n_2)/n_1$ meaning that as the co-ordination numbers are increased the ratio can be held constant so long as their ratio is maintained.
   \par
   As with the previous systems, we can find the behaviour of the location of the initial increase in timescales. Solving for maximal curvature is prohibitive for $n_1\neq n_2$, but the behaviour can be captured through a characterisation of the concentration at which the rate of dissociation doubles from its dilute limit, giving
   \begin{align}
   \frac{C_0}{C_{\rm eff}}&=\frac{K_1K_2(n_1+k_rn_2)}{2k_rC^2_{\rm eff}n_1n_2(n_1-n_2)^2}(\sqrt{\chi^2+\phi}+\chi)+\mathcal{O}(\varepsilon^3)\\
   \chi&={n_1} ({k_r} {n_2}+{n_2}+1)+{k_r} (1-2 {n_2}) {n_2}-2 {n_1}^2\\
   \phi&=8n_1n_2(n_1-n_2)^2,
   \end{align}
   which reduces to Eq.~(\ref{cn1n21}) for $n_2=1$. Keeping $n_1>n_2$ we can expand around $n_2/n_1\sim 0$ giving
   \begin{align}
   \frac{C_0}{C_{\rm eff}}&=\frac{K_1K_2}{C^2_{\rm eff}n_1}+\frac{K_1K_2(1+3n_2)}{2C^2_{\rm eff}n_1^2}\nonumber\\
   &\quad+\frac{K_1K_2(1+n_2(5+2n_2))}{4C^2_{\rm eff}n_1^3}+\mathcal{O}(\varepsilon^3(n_2/n_1)^4)
   \end{align}
   broadly demonstrating that the initial increase in timescale occurs at lower concentrations for higher co-ordination numbers despite constant co-ordination number ratios.
   \par
In the case of equal co-ordination numbers ($n_1=n_2$) the behaviour is qualitatively almost identical to the linear chain, with two quantitative differences i) the rate of dissociation in the most stable regime decreases approximately inversely proportionally to the co-ordination number (Eq.~(\ref{Goffopenn1n2}) and ii) the onset of the responsive regime occur at lower concentrations. By defining such an occurrence as the point the maximal curvature on log-scale, use of the Ansatz $C_0=\alpha K_1^2/C_{\rm eff}^2$ reveals the onset of the responsive regime, in the symmetric case $K_2/K_1=k_r=1$, to occur at approximately 
\begin{align}
\frac{C_0}{C_{\rm eff}}=\frac{K_1^2}{2(n_1-1)C^2_{\rm eff}}+\mathcal{O}(\varepsilon^3).
\end{align}
suggesting sensitivity at lower concentrations for higher co-ordinations numbers.
\par
Finally, we note that for the discussion of percolation in the main text we require a probability of site occupancy which is given by
\begin{align}
P_{\rm occ}&=\frac{2\mathbb{E}[N_c]+\mathbb{E}[n_1^B]+\mathbb{E}[n_2^B]}{n_1+n_2}.
\end{align}
For the $n_1=n_2=4$ square lattice with symmetric kinetics ($K_2=K_1$, $k_r=1$) this is approximated with the solution in Eqs.~(\ref{e12})-(\ref{En}) as
\begin{widetext}
\begin{align}
P_{\rm occ}&=\frac{2 {K_1} \left({K_1}-\sqrt{{C_0^2}+2 {C_0} (6 {C_{\rm eff}}+{K_1})+{K_1^2}}\right)+{C_0^2}+{C_0} (16 {C_{\rm eff}}+3 {K_1})}{{C_0^2}+2 {C_0} (8 {C_{\rm eff}}+{K_1})+{K_1^2}}.
\end{align}
This corresponds to bulk concentration
\begin{align}
C_0&=\frac{-2 \sqrt{16 {C_{\rm eff}}^2 (P_{\rm occ}-1)^2+4 {C_{\rm eff}} {K_1} (P_{\rm occ}-2) (P_{\rm occ}-1)+{K_1}^2}-P_{\rm occ} (8 {C_{\rm eff}}+{K_1})+8 {C_{\rm eff}}+2 {K_1}}{P_{\rm occ}-1}
\end{align}
\end{widetext}
which asymptotically is given by
\begin{align}
\frac{C_0}{C_{\rm eff}}&=\frac{(4-P_{\rm occ})P_{\rm occ}K_1^2}{16C^2_{\rm eff}(1-P_{\rm occ})^2}+\mathcal{O}(\varepsilon^3).
\end{align}
The critical occupation probability found through simulation for the parameters $K_2=K_1=10^{-9}$ and $C_{\rm eff}=10^{-6}$ is $P_{\rm occ}\simeq 0.555$ leading to critical concentration
\begin{align}
\frac{C^{\rm crit}_0}{C_{\rm eff}}&\simeq 0.603\frac{K_1^2}{C^2_{\rm eff}}+\mathcal{O}(\varepsilon^3).
\end{align}
\end{appendix}
\end{document}